\documentclass[11pt]{article}
\usepackage{amssymb}
\usepackage{amsmath}
\usepackage{amsfonts}
\usepackage{stmaryrd}
\usepackage{xypic}
\usepackage[all]{xy}
\usepackage{ulem}
\usepackage{color}
\newcommand{\nc}{\newcommand}
\nc{\rr}{\color{red}}
\nc{\bl}{\color{blue}}
\nc{\gr}{\color{green}}
\usepackage[english]{babel} 
\usepackage[bookmarksnumbered,
plainpages=false,colorlinks=true,linkcolor=black,
citecolor=black,urlcolor=blue]{hyperref}
\newcommand{\beq}{\begin{equation}}
\newcommand{\bee}{\end{equation}}
\newcommand{\beqa}{\begin{eqnarray}}
\newcommand{\eeqa}{\end{eqnarray}}
\newcommand{\noi}{\noindent}
\newcommand{\lb}{\left| \hskip -.2truecm \left|}
\newcommand{\rb}{\right> \hskip -.25truecm \right>}

\makeatletter
\def\eqnarray{\stepcounter{equation}\let\@currentlabel=\theequation
\global\@eqnswtrue
\global\@eqcnt\z@\tabskip\@centering\let\\=\@eqncr
$$\halign to \displaywidth\bgroup\@eqnsel\hskip\@centering
  $\displaystyle\tabskip\z@{##}$&\global\@eqcnt\@ne
  \hfil$\displaystyle{\hbox{}##\hbox{}}$\hfil
  &\global\@eqcnt\tw@ $\displaystyle\tabskip\z@
  {##}$\hfil\tabskip\@centering&\llap{##}\tabskip\z@\cr}
\@addtoreset{equation}{section}
  \def\theequation{\thesection.\arabic{equation}}
\makeatother
\begin{document}

\title{An infinite supermultiplet of massive higher-spin fields}

\author{
{\sf Xavier Bekaert$^a$ \space}
{\sf Michel Rausch de Traubenberg$^b$ \space}
{\sf Mauricio Valenzuela$^a$}\\
{}\\
{\small $^a$ Laboratoire de Math\'ematiques et Physique Th\'eorique}\\
{\small Unit\'e Mixte de Recherche $6083$ du CNRS, F\'ed\'eration Denis Poisson}\\
{\small Universit\'e Fran\c{c}ois Rabelais, Parc de Grandmont}\\
{\small 37200 Tours, France}\\
{\tt \small bekaert@lmpt.univ-tours.fr}, {\tt \small valenzue@lmpt.univ-tours.fr}\\
{}\\
{\small $^b$ IPHC-DRS, UdS, CNRS, IN2P3 }\\
{\small 23 rue du Loess}\\
{\small  67037 Strasbourg Cedex 2, France}\\
{\tt \small Michel.Rausch@IReS.in2p3.fr}
}

\date{
}

\maketitle

\begin{abstract}
The representation theory underlying the infinite-component relativistic wave equation written by Majorana is revisited from a modern perspective. On the one hand, the massless solutions of this equation are shown to form a supermultiplet of the superPoincar\'e algebra with tensorial central charges; it can also be obtained as the infinite spin limit of massive solutions. On the other hand, the Majorana equation is generalized for any space-time dimension and for arbitrary Regge trajectories. Inspired from these results, an infinite supermultiplet of massive fields of all spins and of equal mass is constructed in four dimensions and proved to carry an irreducible representation of the orthosymplectic group $OSp\,(1|4)$ and of the superPoincar\'e group with tensorial charges.
\end{abstract}

\vspace{1cm}

\section{Introduction}
Despite several decades of study, the problem of constructing covariant consistent 
interactions for higher-spin fields (i.e.
spin $s>2$) is still only partially solved, and has turned out to be among the most 
intriguing and challenging problems of field
theory, already at the classical level. In this area of research, it is a 
common place to 
\textit{a posteriori} view string  
theory as a concrete example of such a consistent interacting theory. From a group-theoretical point of view, the spectrum of string 
theory on Minkowski space-time can be described as an infinite sum of unitary irreducible representations (UIRs) of the Poincar\'e group  where all higher-spin representations are massive.
There is an infinite number of fields with increasing masses for any given spin. 
Most studies on higher-spin field theories focus on the truncation of this terribly huge spectrum to the leading Regge trajectory, where each UIR has multiplicity one. Several arguments suggest that such truncations might be consistent, at least in some high-energy regime, and are either looked at as useful toy models or as candidate fundamental theories in themselves.

The birth of higher-spin field quest can be traced back to the early thirties with the pioneering work of Majorana
\cite{Majorana} which, surprisingly enough, remained almost completely unnoticed during three decades, although it anticipated many later developments who received considerable attention from the mathematical physics community: infinite-component relativistic wave equations, UIRs of Lorentz, Poincar\'e and anti de Sitter groups, \textit{etc}.\footnote{A very concise and inspiring account of Majorana's publication itself and of the history of infinite-component wave equations can be found in \cite{Casalbuoni:2006fa}.} The infinite-component wave equation of Majorana was rediscovered independently by Gel'fand and Yaglom \cite{Gelfand} in the late fourties but its genuine revival was due to the efforts of Fradkin \cite{Frad}.
In order to underline the premonitory character of Majorana's ideas, one could mention for instance that the solutions of the linear 
infinite-component relativistic wave equation he proposed not only contain massive UIRs of Poincar\'e group for all spins but also the more exotic ``tachyonic'' and ``continuous-spin'' representations \cite{Sudarshan:1970ss} while it is only\footnote{Esposito and Recami presented several evidences in the research notebooks of Majorana \cite{Recami,Quaderni} supporting the thesis that he might have obtained Dirac-like equations for single massive fields of arbitrary spin even before developing his infinite-component equation. According to Majorana himself, the paper \cite{Majorana} gave only ``a short summary'' of his work on this subject \cite{Leipzig21011933}.} in the late thirties that equations describing a single elementary (massive or massless) particle of any spin were introduced by Dirac, Fierz and Pauli \cite{Fierz}. In the sixties, the proliferation of hadrons with large spin $s$ and mass spectrum roughly described by a linearly rising Regge trajectory 
\begin{equation}
m^2\,=\, \frac{s\,-\,\alpha_0}{\alpha^\prime}\,,
\label{hadron}
\end{equation}
with Regge slope $\alpha^\prime$ and intercept $\alpha_0\,$, was one of the main mystery of strong interaction physics. This prompted an intensive study of various infinite-component relativistic wave equations (sometimes coming 
from the first quantization of some mechanical model) leading to infinite towers of higher-spin particles whose mass is related to the spin. (See \cite{Bohm} and references therein. A concise review of infinite-component relativistic wave equations and dynamical groups can be found in \cite{Barut}.) Unfortunately, the equation of Majorana leads to an unobserved decreasing  Regge trajectory
\begin{equation}
m\,=\,\frac{M}{s+\frac12}
\label{Majoron}
\end{equation}
together with a spectrum of tachyonic particles and continuous-spin massless particles. Analogous problems were shown to automatically arise by Grodsky and Streater \cite{Grodsky:1968zz} for most reasonable avatars of Majorana's seminal work. Their no-go theorem undermined the corresponding programmes of research (dynamical groups and current algebras) while two distinct (parton \textit{versus} dual) models of hadrons started to attract attention (and gave rise, respectively, to quantum chromodynamics and string theory).

Nevertheless, the representation theory behind Majorana construction is re-examined here from a contemporary perspective because many of its key ingredients, such as the singletons, play now a prominent role in the non-Abelian massless higher-spin theory on anti de Sitter space time (see \textit{e.g.} \cite{Vasiliev:2004qz,BCIV} for some reviews). The paper \cite{Majorana} is usually referred to as the first appearance in elementary particle physics of unitary representations of $Spin(1,3)\cong SL(2,{\mathbb C})$, the double cover of the Lorentz group, but it is almost never mentioned that actually Majorana also presented and made a decisive use of two unitary representations of the bigger group $Spin(2,3)\cong Sp(4,{\mathbb R})$, the double covering of the anti de Sitter isometry group. Since this modern point of view is rather anachronistic, the paper \cite{Dirac:1963ta} of Dirac on these ``remarkable'' representations is traditionally referred to as the seminal paper on th
 e 
``singleton'' representations (using the terminology introduced by Flato and Fr\o nsdal \cite{Flato:1978qz} much later).
Still, this surprising appearance of two \textit{a priori} unrelated structures in the same context is very suggestive, and so it is not excluded that Majorana-like constructions could play a role in the mysterious spontaneous symmetry breaking of higher-spin gauge symmetries. This possibility motivates a thorough examination of a class of Majorana-like infinite-component wave equations from a contemporary perspective.\footnote{Somewhat similarly, an old mechanical model (inspired by the string model of hadronic physics) producing an infinite tower of massless and massive particles has been revisited very recently \cite{Bengtsson:2009nk} in the modern light of the interaction problem for higher-spin gauge theories.} 

\subsection{Summary of the main results}

In the present paper, the generation of various infinite spectrums of masses for higher-spin particles from a single relativistic wave equation but with an infinite number of components is investigated. The analysis is essentially restricted to four-dimensional Minkowski space-time but most of the results allow straightforward higher-dimensional extensions.

The focus is put here on the intermediate situation where the generated mass spectrum is neither rising like (\ref{hadron}) nor decreasing like (\ref{Majoron}) but is instead a ``horizontal'' Regge trajectory. In other words, there is only one mass-shell but of infinite degeneracy\footnote{Notice that this property is in agreement with the main conclusion of Grodsky-Streater's no-go theorem \cite{Grodsky:1968zz}. In the present context, this degeneracy is not considered as a fatal disease but, on the contrary, as a natural feature of a massive higher-spin multiplet.}: $m^2=$ constant for all (integer and half-odd-integer) spin $s\,$.
 The proposed supermultiplet is thus an infinite tower of particles of equal mass but with all spins. This collection of particles is shown to carry an irreducible representation of the $\mathfrak{osp}(1|4)$ superalgebra. More concretely, the physical components of each massive field is a spin-$s$ representation and the direct sum of all such representations (with multiplicity one) precisely fits in the UIR of $\mathfrak{osp}(1|4)$ called the ``singleton supermultiplet'' (see \textit{e.g.} \cite{Nicolai:1984hb} for a pedagogical introduction). This argument is valid for any fixed plane wave so that $OSp(1|4)$ symmetry group commutes with the space-time translation group.
  From the space-time point of view, the supermultiplet proposed here carries also a representation of the superPoincar\'e algebra with tensorial ``central'' charges. This property is reminiscent of the supersymmetric particle models with tensorial central charges \cite{Bandos:1999qf} producing upon quantization the infinite supermultiplet of massless particles with all spins given in \cite{Fronsdal:1985pd}. This model is of direct relevance \cite{Plyushchay:2003gv} in the non-Abelian higher-spin gauge theory  on $AdS_4\,$. However, it should be stressed that our supermultiplet is not a usual one in the sense that the corresponding ``translation'' operators, mix space-time translation and ``spinning'' degrees of freedom. The usual translation of the superalgebra are recovered for the massless solution of the Majorana equation, or equivalently, in the continuous-spin limit of the massive supermultiplet.

This limit deserves several comments because it is of interest in itself. 
In the recent paper \cite{Horvathy:2007pm}, linear relativistic wave equations unifying the spin-0 equation introduced by Dirac in \cite{Dirac} and its spin-1/2 counterpart \cite{Stau2} were proposed and generalized (by means of the Majorana equation) to a supersymmetric theory of massive higher-spin particles. The resulting theory is characterized by a nonlinear symmetry superalgebra that, in the infinite-spin limit, reduces to the super-Poincar\'e algebra with or without tensorial central charge. This subtle infinite-spin ($s\rightarrow \infty$) and zero-mass ($m\rightarrow 0$) limit of massive higher-spin representations with the product $ms=M$ kept fixed has actually been studied recently \cite{Bekaert:2005in}
and leads to one of the two (either bosonic or fermionic) ``continuous-spin'' representations of the Poincar\'e group (whether the spins $s$ are all integer or all half-odd-integer). Actually, the Majorana equation provides a particular realization of this limit because the massless sector can be understood as the infinite-spin limit of a couple of particles of spins $s+1/2$ and $s$ with masses determined by the Regge trajectory (\ref{Majoron}). Notice that the ratio of their masses and number of degrees of freedom goes to unity in the limit, as it should be for any exact supermultiplet. After the limit, the couple of (bosonic and fermionic) continuous-spin particles forms a supermultiplet: the massless sector of Majorana's equation is supersymmetric, a surprising fact which seems to have been unobserved previously. 
A continuous-spin supermultiplet was already found in \cite{Brink:2002zx} but non-trivial central extensions were not found. Therefore, one may identify the superPoincar\'e algebra obtained in \cite{Horvathy:2007pm} and in the present paper as a possible extensions of \cite{Brink:2002zx} with tensorial central charges.

\subsection{Structure of the paper}

 The plan is as follows: Section \ref{anydim} is a review of infinite-component Dirac-like equations of the type introduced by Majorana. These equations are introduced from a very general perspective which allows their straightforward generalization to other representations of the Lorentz algebra and to any dimension. The Di and Rac representations are instrumental in Majorana's construction, so the section \ref{sec-osp} is devoted to a detailed review of the many aspects of these representations in their (probably most convenient) realization in terms of Fock space.
This allows to review briefly in Section \ref{Majsp} the spectrum of particles for the infinite-component Majorana equation. More generally, infinite-component relativistic wave equations whose spectrum of solutions provide arbitrary Regge trajectories are presented in Section \ref{Reggelike}.
A dictionary between the planar harmonic oscillator states and the infinite collection of massive particles in rest frame of all spins and of equal mass is provided in Section \ref{harmonic}.
Following the philosophy of dynamical groups \cite{Bohm,Barut}, we boost this infinite collection of states and prove in Section \ref{S-P sec} that it forms an infinite supermultiplet: it
spans an irreducible representation of the orthosymplectic group $OSp\,(1|4)$ 
and of the superPoincar\'e group with tensorial central charges. The latter groups respectively correspond to ``spinning'' versus ``space-time'' symmetries. The section \ref{concl} is the conclusion, where in particular, some aspects about (super)symmetry breaking are briefly discussed. The main ingredient for building the various representations considered is the Weyl algebra $A_2\,$, so an appendix is devoted to several of its finite-dimensional algebras which are used here.

\section{Infinite-component Dirac-like equations in any dimension}\label{anydim}

In order to stress the degree of generality of the underlying philosophy behind the infinite-component Majorana equation, this section proposes some possible generalisations in the light of modern knowledge of representation theory.\footnote{An exhaustive treatment of \textit{finite-component} Dirac-like equations in $D=4$ dimensions was performed by Bhabha in \cite{Bhabha:1945zz}.} One of the key idea of Majorana was to write down a linear wave equation which formally resembles to Dirac's one except that the wave function takes value in a unitary Lorentz algebra module $V$ (or ``representation space''), so that the Hilbert space $\cal H$ of solutions is a \textit{reducible} Poincar\'e algebra module decomposing
as an \textit{infinite} sum of irreducible Poincar\'e algebra modules where the values of the quadratic (momentum squared) and quartic (Pauli-Lubanski vector squared) Casimir operators are related by means of the wave equation. In other words, the mass spectrum of particles is related to the spin.\footnote{Actually, Majorana's motivation was not to generate a mass spectrum but instead the property that the equation admits only positive energy solutions.} 
As stressed here such a construction can of course be done in any space-time dimension and by using various unitary representations of the Lorentz group.
The restriction to $D=4$ dimensions and Majorana representation of the Lorentz group can be understood as a very particular case.  (For a review of the UIRs of Poincar\'e groups and of their correspondence with relativistic wave equations in any space-time dimension $D\geqslant 4\,$, see \textit{e.g.} \cite{Bekaert:2006py}.)

Concretely, let $\psi$ be a wave function taking values in some Lorentz $\mathfrak{so}(1,D-1)$-module $V$ where the generators of $\mathfrak{so}(1,D-1)$ are realised as Hermitian operators $S_{\mu\nu}$, acting on the spinning degrees of freedom (or ``components'') that span the module $V$, and satisfying the commutation relations
\begin{equation}
[S_{\mu\nu},S_{\lambda\rho}]=i(\eta_{\mu\lambda}S_{\nu\rho}+\eta_{\nu\rho}S_{\mu\lambda}-
\eta_{\mu\rho}S_{\nu\lambda}-\eta_{\nu\lambda}S_{\mu\rho})
\label{AdS0}
\end{equation}

\noi
where $\eta_{\mu \nu}=\text{diag}(-1,1,\cdots,1).$
Consider the generators 
of the Poincar\'e algebra,
\begin{equation}\label{poincare}
P_\mu=-i\partial/\partial x^\mu,\qquad J_{\mu\nu}=x_{\mu}P_\nu-x_{\nu}P_\mu+S_{\mu\nu}, \qquad
\mu,\nu=0,1,...,D-1
\end{equation}
where $x^\mu$ are the space-time coordinates, $P_\mu$ their conjugated momenta and $ S_{\mu\nu}$ is now interpreted as
the spin part of the Lorentz generator $J_{\mu\nu}\,$. 

 As it is, the space of wave functions $\psi$ is by construction a module of the Poincar\'e algebra, and thus, of its Lorentz subalgebra generated by $J_{\mu\nu}$. Notice that even when the $\mathfrak{so}(1,D-1)$-module $V$ generated by $S_{\mu\nu}$ is irreducible, it does not implies the irreducibility of $\mathfrak{so}(1,D-1)$-module generated by $J_{\mu\nu}$.
 The important point we would like to emphasize, is that it is necessary to have some relativistic wave equations (\textit{e.g.}  the Dirac, Proca equation, etc.) in order to determine what is the physical content of the theory and, for instance,  determine whether the representation of the Poincar\'e algebra is irreducible and unitary. Indeed, the relevant $\mathfrak{iso}(1,D-1)$-module is the submodule $\cal H$ of solutions of the relativistic wave equations. This subtlety is well-known but may be sometimes confusing. For instance, we will stress that the infinite-component wave function of Majorana takes values in a UIR of the Lorentz algebra but the Hilbert space of solutions of this wave equation carries a \textit{reducible} representation of the Poincar\'e algebra.

Till now, the discussion has been completely generic so let us focus on the general recipe for preparing a Majorana-like infinite-component relativistic wave equation goes as follows:
\begin{enumerate}
	\item write down a Dirac-like equation
\begin{equation}
\Big(P_\mu\Gamma^\mu-M\Big)\psi=0\,,
\label{Diraceq}
\end{equation}
where $M$ is a non-vanishing parameter, say positive $M>0\,$, with the dimension of a mass;
 \item impose that the ``Gamma matrices'' $\Gamma^\mu$ transform as
 vectors under the adjoint action of Lorentz algebra: 
\begin{equation}
[S_{\mu\nu},\Gamma_\lambda]=i(\eta_{\mu\lambda}\Gamma_\nu-\eta_{\nu\lambda}\Gamma_\mu)\,,
\label{AdS1}
\end{equation}
in order to ensure Lorentz invariance;
  \item require that the components of the wave function transform in a representation of the Lorentz algebra whose generators are proportional to the commutator of ``Gamma matrices'' via the usual relation
\begin{equation}
S_{\mu\nu}:=-i\,[\Gamma_\mu,\Gamma_\nu]\,;
\label{AdS2}
\end{equation}
  \item relax Dirac's assumption that (\ref{Diraceq}) 
implies the Klein-Gordon equation $P^2=-M^2\,$; 
  \item assume that the components of the wave function span some \textit{unitary} (hence infinite-dimensional) module $V$ of the pseudo-orthogonal algebra $\mathfrak{so}(2,D-1)$ spanned by $\Gamma_\mu$ and $S_{\nu\rho}\,$.
\end{enumerate}
Physically speaking, the fourth assumption is necessary if one wants to generate a non-trivial mass spectrum. Mathematically speaking, it means that the $\Gamma$'s are not assumed to span a Clifford algebra. Actually, no hypothesis is made on their \textit{anti}commutator. However, the assumptions (\ref{AdS0}), (\ref{AdS1}) and (\ref{AdS2}) state that $\Gamma_\mu$ and $S_{\nu\rho}$ span together a representation of the pseudo-orthogonal algebra $\mathfrak{so}(2,D-1)$. This lead to the last assumption, which implies that $V$ is also a unitary (maybe reducible) module of the Lorentz subalgebra $\mathfrak{so}(1,D-1)\subset\mathfrak{so}(2,D-1)\,$. The main differences between Majorana infinite-component and Dirac finite-component equation are that,
for  the former, the wave function transforms in a unitary representation and that the $\Gamma^\mu$ are all Hermitian (in such a way that their spectrum is real), which is impossible for Clifford algebras with Lorentzian signature.\footnote{
 Indeed, more generally,  the second and third assumptions above
are  very closely related to the parafermions 
\cite{para,para1}.
The difference being that he
considered order $p$ parafermions, that is a finite dimensional representation
of the Lorentz algebra. Order one parafermions correspond to the
Clifford algebra, order two parafermions to the Kemmer-Duffin-Petiau algebra, \textit{etc} \cite{para}.} 
Notice that the gender of the $D$th extra direction associated to the generator $S_{\mu,D}:=\Gamma_\mu$ is fixed by the relative sign in the commutation relation (\ref{AdS2}), so another possibility would correspond to the commutation relations of the de Sitter algebra $\mathfrak{so}(1,D)\,$. This choice is rejected because there do not exist any UIR of $\mathfrak{so}(1,D)$ such that $S_{0,D}:=\Gamma_0$ is positive-definite.  In other words, if the generators of $\mathfrak{so}(1,D)$ are all Hermitian then the spectrum of $\Gamma_0$ is, of course, real but it automatically contains negative eigenvalues.

The quadratic Casimir operator of the
Lorentz subalgebra $\mathfrak{so}(1,D-1)$ is the square of the
generators $S_{\mu\nu}$:
\begin{equation}{\cal C}_2\Big(\mathfrak{so}(1,D-1)\Big) \,=\,
\frac12\,J^{\mu\nu}J_{\mu\nu}\,.
\label{BeBo:QuadratiC}
\end{equation}
The quadratic Casimir
operator of the Poincar\'e algebra $\mathfrak{iso}(1,D-1):={\mathbb
R}^{D-1,1}\niplus\mathfrak{so}(1,D-1)\,$ is the
square of the momentum
\begin{equation}
{\cal C}_2\Big(\mathfrak{iso}(1,D-1)\Big) \,=\,
-P^{\mu}P_{\mu}\,, 
\label{BeBo:quadratiC}\end{equation} while the
quartic Casimir operator is
\begin{equation}
{\cal C}_4\Big(\mathfrak{iso}(1,D-1)\Big) \,=\, -{1\over 2}P^2J_{\mu\nu}J^{\mu\nu}
+J_{\mu\rho}P^\rho J^{\mu\sigma}P_\sigma\,,
\label{BeBo:Quartic}
\end{equation}
which, for $D=4$,
is the square of the Pauli-Lubanski vector $W^{\mu}$, 
\begin{equation}
W^{\mu}:=\frac{1}{2}\,\varepsilon^{\mu\nu\rho\sigma}
J_{\nu\rho}P_{\sigma}\,,
\label{PLub}
\end{equation}
(where $\varepsilon^{0123}=1$).
Notice that the quadratic and quartic Casimir operators essentially classify
the UIRs in $D=4\,$, but this is no more true in higher dimensions where more
Casimir operators are necessary. Moreover, one should stress that the eigenvalues of the Casimir operators do not
characterize uniquely an irreducible representation (for instance,
the quadratic and quartic Casimir operators vanish for all
helicity representations).

The sign of the quadratic Casimir operator (\ref{BeBo:quadratiC}) of the Poincar\'e algebra determines the gender of the momentum $P_\mu\neq 0$ (when $M\neq 0$) and thereby its stabilizer (or ``little'') algebra 
$\mathfrak{l}_{\,\mbox{sgn}({\cal C}_2)}$
depending on $\mbox{sgn}(P^2)=-\mbox{sgn}({\cal C}_2)\,$, the sign  of the momentum square. This implies that the unitary Poincar\'e module $\cal H$ of solutions $\psi$ to the Dirac-like equation (\ref{Diraceq}) admits the obvious decomposition in submodules
$$
{\cal H}={\cal H}_{-}\oplus {\cal H}_{0}\oplus{\cal H}_{+}\,,
$$
where the direct sum is actually over the sign of the momentum square, $\mbox{sgn}(-P^2)\in\{-1,0,+1\}\,$.
The quartic Casimir operator (\ref{BeBo:Quartic}) of the Poincar\'e algebra
can be evaluated in components in the canonical frame adapted to the given momentum.
These relation between the various values of the Casimir operators can be summarized in the following table (see \cite{Bekaert:2006py} for details).
\vspace{5mm}
\begin{table}[ht]
\caption{}
\begin{center}
\begin{tabular}{|c|c|c|c|}
 \hline
  Quadratic Casimir & Stability algebra & UIR & Quartic Casimir operator \\\hline\hline
  $>0$ & $\mathfrak{l}_+=\mathfrak{so}(D-1)$ & \small{Massive} & ${\cal
C}_2\Big(\mathfrak{iso}(1,D-1)\Big)\times{\cal C}_2\Big(\mathfrak{so}(D-1)\Big)$\\\hline
  $=0$ & $\mathfrak{l}_0=\mathfrak{iso}(D-2)$ & \small{Massless} & ${\cal
C}_2\Big(\mathfrak{iso}(D-2)\Big)$\\\hline
  $<0$ & $\mathfrak{l}_-=\mathfrak{so}(1,D-2)$  & \small{Tachyonic} & ${\cal
C}_2\Big(\mathfrak{iso}(1,D-1)\Big)\times{\cal
C}_2\Big(\mathfrak{so}(1,D-2)\Big)$ \\\hline
\end{tabular}
\end{center}
\end{table}
\vspace{5mm}
The important lesson is that for given values of the quadratic Casimir operators of the Poincar\'e group \textit{and} of the little group, the quartic Casimir operator is completely determined. 
Therefore it is natural to decompose the $\mathfrak{so}(1,D-1)$-module $V$ 
as a direct sum of irreducible $\mathfrak{l}_\epsilon$-submodules $V_\epsilon^J$ labelled by the index $J$ 
for fixed $\epsilon\in\{-1,0,+1\}\,$: 
$V=\oplus_J\, V_\epsilon^J\,$. This decomposition can be computed via the known branching rules for the restriction of $\mathfrak{so}(1,D-1)$ to its subalgebra $\mathfrak{l}_\epsilon$\,.
From Wigner's method of induced representations, one expects
that, in each Hilbert subspace ${\cal H}_\epsilon$ 
of solutions there is a one-to-one correspondence between any irreducible 
$\mathfrak{l}_\epsilon$-submodules $V_\epsilon^J$ and an irreducible $\mathfrak{iso}(1,D-1)$-submodule ${\cal H}_\epsilon^J\,$. Therefore, the Hilbert space of solutions decomposes into irreducible Poincar\'e-modules as follows
$$
{\cal H}\,=\bigoplus\limits_{\epsilon\in\{-1,0,+1\}}\bigoplus\limits_J\,{\cal H}^J_\epsilon\,.
$$
A particular example might clarify these last steps. Let us take as unitary irreducible $\mathfrak{so}(2,D-1)$-module $V$ the ``conformal scalar field'' 
on {${\mathbb R}^{1,D-2}$} space-time for $D\geqslant 4$, which is  denoted by ${\cal D}(\frac{D-3}2,0)$ in the literature (see \textit{e.g.} \cite{Engquist:2007vj} for a short review). Let us consider a plane wave in the massive sector ${\cal H}_+$ of solutions of the Dirac-like equation (\ref{Diraceq}). In the rest frame, the momentum takes the simple form $P^\mu=(m,0,\ldots,0)$ hence the eigenvalue of the Dirac-like operator is equal to $m\,\Gamma_0=M\,$. Thus the massive spectrum is entirely determined by the spectrum of the operator $\Gamma_0\,$, which is interpreted as the ``energy'' when the algebra $\mathfrak{so}(2,D-1)$ is interpreted as the anti de Sitter isometry algebra. The rotation algebra $\mathfrak{l}_+\cong \mathfrak{so}(D-1)\subset \mathfrak{so}(2,D-1)$ characterizes the massive representations. Then, $V={\cal D}(\frac{D-3}2,0)$ can be 
decomposed as the direct sum of irreducible $\mathfrak{so}(D-1)$-modules ${\cal D}_J$ (labeled by a Young diagram made of a single row of $J$ boxes; they generalize the ``spin-$J$'' $\mathfrak{so}(3)$-modules) as follows:
$$
{\cal D}\Big(\frac{D-3}2,0\Big)=\bigoplus\limits_{J\in\mathbb N} {\cal D}_J\,,
$$
where each irreducible $\mathfrak{so}(D-1)$-module $V_+^J={\cal D}_J$ is also an eigenspace of $\Gamma_0$:
$$
\Gamma_0\,{\cal D}_J=\Big(J+\frac{D-3}2\Big){\cal D}_J\,.
$$
This shows that the massive sector ${\cal H}_+$ of solutions of the Dirac-like equation is the infinite direct sum of irreducible $\mathfrak{iso}(1,D-1)$-modules ${\cal H}^J_+$ describing a particle of ``spin'' $J$ and mass
\begin{equation}\label{Majorongen}
m_J=\frac{M}{J+\frac{D-3}{2}}\,,\qquad J=0,1,2,...,  
\end{equation}
generalizing the formula (\ref{Majoron}) for bosons.  This shows explicitly that though the wave function spans an irreducible Lorentz-module, the corresponding space of solution of the wave equation is reducible. The same result should apply to the ``conformal spinor field,'' \textit{i.e.} for half-odd-integers $J$.

It is worth mentioning that the formula (\ref{Majorongen}) applies also in the three-dimensional case, $m_J=M/J$, but the spin $J$ is not quantized: it may take continuous real values. In fact, the Majorana equation in $D=3$ even fixes the spin $J\in\mathbb R$ and describes anyons (``fractional spin particles''), see \textit{e.g.} \cite{Jackiw,Horvathy:2006pw} and reference therein. 
The main reason is that the Dirac-like operator $P\cdot\Gamma=\frac{1}{2} \epsilon_{\mu\nu\lambda}P^\mu J^{\nu\lambda}$ is a quadratic Casimir operator of the Poincar\'e algebra $\mathfrak{iso}(1,2)$ since in $D=3$ the $\Gamma_\mu$ operator is equivalent to the dual of Lorentz transformation, \textit{i.e.} $\Gamma_\mu=\frac{1}{2} \epsilon_{\mu\nu\lambda}S^{\nu\lambda}$ implies (\ref{AdS2}).

Majorana's work corresponds to the particular case where $D=4$ and the unitary module of the $AdS_4$ algebra is the singleton supermultiplet, \textit{i.e.} the sum of the ``Di'' and ``Rac'' UIRs (following the terminology introduced in \cite{Flato:1978qz}) or, rephrasing it, in the trivial frame ($P^\mu=0$) the Majorana wave function is equivalent to singleton supermultiplet. These two ``remarkable'' UIRs of $\mathfrak{so}(2,3)$
share the property that their restriction to $\mathfrak{so}(1,3)$ 
remains irreducible  (this, and previously mentioned facts on the $AdS_4$ algebra are presented in more details for the $D=4$ case, in Appendix \ref{app}.)

\section{The $OSp(1|4)$ Di-Rac-Majorana representation}\label{sec-osp}

The representation of $\mathfrak{so}(2,3)$ employed by Majorana in the Dirac-like equation (\ref{Diraceq}) is very exceptional. He looked for some unitary representations of $\mathfrak{so}(2,3)$ in order to get positive-energy solutions from his  
equation, as explained in the previous section.
The isomorphism of algebras $\mathfrak{sp}(4)\cong \mathfrak{so}(2,3)$ allows to construct representations of the Anti de Sitter algebra as symmetrized quadratic products in the Hermitian generators of the Weyl algebra $A_2$
(realized here by the operators $q_i$ and $\eta_i=-i\partial/ \partial q_i,\,i=1,2\,$;
see the appendix \ref{app} for more details). 
In fact, with this procedure Majorana implicitly introduced 
an additional structure for the spinning degrees of freedom which in turn is related to
the metaplectic group $Mp(4):=\widetilde{Sp}(4)\,$, the double cover of the symplectic group $Sp(4)\,$. 
The Fock spaces were rigorously introduced as modules of the groups $Mp(2n)$ by
Weil and Shale in mathematics under the name of metaplectic modules, but they were 
known already to physicists since they underlie the harmonic oscillator with $n$ degrees of freedom.
The metaplectic (or Weil) representation is a faithful unitary representation of the metaplectic group $Mp(4)\,$,
and it is only a projective (``double-valued'') representation of the symplectic group $Sp(4)\,$.
The metaplectic group $Mp(4)$ is not a matrix group: it has no faithful finite-dimensional representations.
The metaplectic representation is reducible: the metaplectic module $\cal M$ decomposes in two irreducible $Mp(4)$-modules,  say ${\cal M}={\cal M}_+ \oplus{\cal M}_-\,$. Actually, the Fock space $\cal M$ is ${\mathbb Z}_2$-graded by the parity of the number operator. Surprisingly, the Di and Rac UIRs of $Sp(4)\cong Spin(2,3)$ are identified with the previous modules for $Mp(4)\,$: Di$\,=\,{\cal M}_-$ and Rac$\,=\,{\cal M}_+\,$. The Di and Rac representations carry half-odd-integer and integer spin respectively, and they can be joined in a representation of $OSp(1|4)$ with the supercharges interchanging the Di and Rac modules.
(See the appendix \ref{app}.)
In the metaplectic representation the operators interchanging the ${\cal M}_\pm$ modules
are naturally given in terms of the generators
of $A_2\,$. However, a subtle but important observation is in order:
The usual supercharges of $OSp(1|4)$ are Grassmann-odd (fermionic), hence they have finite-dimensional representations. Instead, the supercharges of $OSp(1|4)$ are represented here by generators of $A_2$, which are Grassmann-even 
(denominated as well ``bosonized supersymmetry'' \cite{Horvathy:2006pw,BosonSUSY}), therefore we must get an infinite-dimensional representation. We will take full advantage of this fact to get an infinite (massive) super-multiplet.

Let us study these aspects in detail. The generators of the Weyl algebra $A_2$ can be arranged in the vector 
$L_a=(q_1,q_2,\eta_1,\eta_2)$, where $a=1,2,3,4\,$. Defining $M_{ab}:= \frac{1}{2}\{L_a,L_b\}$ we obtain the $\mathfrak{osp}(1|4)$ (anti)commutation relation
\begin{eqnarray}
\begin{array}{l}
[ M_{ab}, M_{cd} ]=i(C_{ac} M_{bd}+C_{bd}M_{ac}+C_{ad}M_{bc}+C_{bc}M_{ad})\\[10pt] 
M_{ab}= \frac{1}{2}\{L_a,L_b\},\qquad [M_{ab},L_c]=i(C_{ac}L_b+C_{bc}L_a),
\end{array}\label{osp-spinor}
\end{eqnarray}
derived from the canonical commutation relations of the Weyl algebra generators,
\begin{eqnarray}\label{A2}
[L_a,L_b]=iC_{ab}, \qquad C_{ab}=
\left(%
\begin{array}{cc}
  0 & I_{2\times2} \\
  -I_{2\times2} & 0 \\
\end{array}
\label{Cab} \right).
\end{eqnarray}
 The isomorphism $\mathfrak{sp}(4)\cong\mathfrak{so}(2,3)$ implies that the symplectic index $a$ is actually also spinorial in the sense that the vector $L_a$ can also be interpreted as a Grassmann-even real spinor (``twistor'') and the antisymmetric (symplectic) matrix $C_{ab}$ also works as a spinor metric (see e.g. \cite{Fronsdal:1985pd})
raising and lowering spinor-symplectic indices as $A^a=A_bC^{ba}$ and $A_a=C_{ab}L^b$, where $C^{ab}=C_{ab}$, $C_{ac}C^{bc}=\delta_a^b$. We are  now able to obtain in a space-time covariant way the singleton representation of the algebra $\mathfrak{so}(2,3)$\footnote{For further details of this representation see appendix \ref{app}, and alternatively ref. \cite{Horvathy:2007pm}.}
\begin{equation}\label{so32g}
S^{\mu\nu}=-\frac{i}{2}(\gamma^{\mu \nu})^{ab} M_{ab},\qquad
\Gamma^\mu=-\frac{1}{4}(\gamma^\mu)^{ab} M_{ab},
\end{equation}
where the Dirac $\gamma-$matrices are taken in the Majorana representation \eqref{gammas}.
As a direct consequence of \eqref{A2} we have the following identities 
\begin{eqnarray}\label{id}
\Gamma^\mu\Gamma_\mu=\frac {1}{2},\qquad \Gamma^\mu
S_{\mu\nu}=S_{\nu\mu}\Gamma^\mu=-\frac{3i}{2}\Gamma_\nu,\qquad
\epsilon^{\mu\nu\lambda\rho}S_{\nu\lambda}\Gamma_\rho=0. 
\end{eqnarray}
It has to be stressed however that the  $\Gamma$ operators do not produce the
 Clifford algebra but instead $\Gamma_\mu\Gamma_\nu=\frac{1}{2}\eta_{\mu\nu}-\frac{3i}{2}S_{\mu\nu}-S_{\mu\lambda}S^\lambda{}_\nu.$ From these relations, the two Casimir operators of $\mathfrak{so}(2,3)$ are easily obtained,
$$
{\cal C}_2\big(\mathfrak{so}(2,3)\big):=\frac12\, S^{AB}S_{AB}=-\frac{5}{4},\qquad {\cal C}^\prime_2\big(\mathfrak{so}(2,3)\big):=V^A V_A=0,\qquad  A,B=0,1,2,3,4,
$$
where $S^{4\mu}:=\Gamma^\mu$ has been defined together with the operator $V^A:=\epsilon^{ABCDE}S _{BC}S_{DE}$ which is identically zero $V^A\equiv 0$. The Levi-Civita tensor $\epsilon^{01234}=1$ and the metric is taken to be 
 $\eta^{AB}=\text{diag}(-1,1,1,1,-1)\,$. Instead, for the Lorentz subalgebra $\mathfrak{so}(3,1)$, the Casimir operator are given by 
\begin{equation}
\label{Casimir-Poin}
{\cal C}_2\big(\mathfrak{so}(1,3)\big):=\frac12\,S^{\mu\nu}S_{\mu\nu}=-\frac{3}{4}, \qquad
{\cal C}^\prime_2\big(\mathfrak{so}(1,3)\big):=\frac14\, \epsilon^{\mu\nu\lambda\rho}S_{\mu\nu}S_{\lambda\rho}=0.
\end{equation}

\noindent
It is convenient to write the $\mathfrak{osp}(1|4)$ commutation relations (\ref{osp-spinor}) in terms of the space-time covariant generators (\ref{so32g}),
\begin{eqnarray}\label{AdSext}
\begin{array}{c}
[S_{\mu\nu},S_{\lambda\rho}]=i(\eta_{\mu\lambda}S_{\nu\rho}+\eta_{\nu\rho}
S_{\mu\lambda}-
\eta_{\mu\rho}S_{\nu\lambda}-\eta_{\nu\lambda}S_{\mu\rho})\\[8pt]
[S_{\mu\nu},\Gamma_\lambda]=i(\eta_{\mu\lambda}\Gamma_\nu-\eta_{\nu\lambda}\Gamma_\mu),
\qquad [\Gamma_\mu,\Gamma_\nu]=-iS_{\mu\nu},\\[8pt]
[S_{\mu\nu},L_a]=-(\gamma_{\mu\nu})_a\,^bL_b,\qquad[\Gamma_\mu,L_a]=\frac{i}{2}(\gamma_\mu)_a\,^bL_b,
\\[8pt]
\{L_a,L_b\}=-2(iS_{\mu\nu}\gamma^{\mu\nu}-\Gamma_\mu\gamma^{\mu})_{ab}.\label{S,L}
\end{array}
\end{eqnarray}
The supersymmetric structure reveals that the $\mathfrak{so}(2,3)$ representation is reducible since it contains a bosonic and a fermionic submodule. This is also reflected in the existence of a nontrivial central element commuting with $\mathfrak{so}(2,3)$. It is the parity operator in the {\it ``spinning'' phase space}, denoted ${\cal R}$, which maps $(q_i,\eta_i)$
to $(-q_i,-\eta_i)$. This operator is unitary since it is associated to a  
$U(1)\subset Sp(4)$ transformation.
Indeed, we have
\begin{equation}\label{RR'}
{\cal R}=-\exp(i2\pi \Gamma_0)=\exp(i2\pi S_{12}).
\end{equation}
This means that ${\cal R}$ can be understood as  
a $2\pi$ rotation in a time-time like or a space-space like plane but by 
a $\pi$ rotation in the plane $\mathbb{R}^2$ with coordinates $(q_1,q_2)\,$. Observe that 
\begin{equation}\label{RLR}
{\cal R}L_a{\cal R}^\dagger=-L_a,\qquad {\cal R}S_{\mu\nu}{\cal
R}^\dagger=S_{\mu\nu},\qquad {\cal R}S_{\mu\nu}{\cal
R}^\dagger=S_{\mu\nu}.
\end{equation}
 Taking into account the equations ${\cal R}^\dagger={\cal R}$, ${\cal R}^2=1$ and (\ref{RLR})
the  following (anti)commutation relations,
\begin{equation}\label{grad}
[{\cal R},S_{\mu\nu}]=0, \qquad [{\cal R},\Gamma_\mu]=0,\qquad
\{{\cal R},L_a\}=0\,,
\end{equation}
are obtained. Thus ${\cal R}$ generates a representation of $\mathbb{Z}_2$ such that
 $S_{\mu\nu}$, $\Gamma_\mu$ are even and $L_a$ are odd operators, 
\textit{i.e.} ${\cal R}$ is the grading operator of the $\mathfrak{osp}(1|4)$ superalgebra (or more generally of $A_2$) 
in the representation (\ref{AdSext}).\footnote{
The ${\cal R}$ operator employed here is, a generalization of the parity operator of the one-dimensional harmonic oscillator previously introduced in \cite{Wigner1950}, and frequently called ``Klein'' or ``reflection'' operator (see \textit{e.g.} \cite{Horvathy:2006pw,Vasiliev:1989qh,Plyushchay:1997ty}).}.
The eigenspaces, ${\cal M}_\pm=\pm {\cal R}{\cal
M}_\pm,$ are invariant modules of $Sp(4)$. We can extract out irreducible representation of $Sp(4)$ introducing of the projector operators,
\begin{eqnarray}\label{projectors}
\Pi_\pm=\frac{1}{2}(1\pm{\cal R}),
\qquad\left(\Pi_\pm\right)^2=\Pi_\pm,\qquad \Pi_+\Pi_-=0, \qquad
\Pi_++\Pi_-=1,
\end{eqnarray}
such that $\Pi_\pm {\cal M}={\cal M}_\pm.$ From (\ref{grad}) and (\ref{projectors}), the irreducible representation of $Sp(4)$ upon these two modules are labelled by $Sp_\pm(4)$ and are generated by, 
\begin{equation}
 \mathfrak{sp}_\pm(4)=\{S_{\mu\nu}^\pm := S_{\mu\nu}\Pi_\pm,\quad \Gamma_\mu^\pm :=
\Gamma_\mu \Pi_\pm\},
\end{equation}
with the algebras $\mathfrak{sp}_\pm(4)$ commuting with each other. The $\mathfrak{sp}_+(4)$ algebra generates the $SO(3,2)$ group and $\mathfrak{sp}_-(4)$ its double cover $Sp(4)$, associated respectively with bosons and fermions, and both are merged in the metaplectic group. 
 Of course, at the level of algebras $\mathfrak{mp}(4) \cong \mathfrak{sp}(4) \cong \mathfrak{so}(2,3)$.
The module $\cal{M}$ is described in Appendix \ref{app} via the
reduction of $\mathfrak{mp}(4)$ under its maximal compact subalgebra, say
$\mathfrak{so}(2) \oplus \mathfrak{so}(3)$. The modules ${\cal M}_\pm$ are then identified
as the lowest-weight modules for the two Cartan generators: the energy $\Gamma_0$ and the spin of the $\mathfrak{so}(3)$
subalgebra. The modules ${\cal M}_-$ and ${\cal M}_+$ are thereby denoted, respectively, as ${\cal D}(\,1/2\,,\,0\,)$ and ${\cal D}(\,1\,,\,1/2\,)$ (see \textit{e.g.} \cite{Nicolai:1984hb} for a review).

The ubiquitous appearance of the above-mentioned modules explains the large number of synonymous terminology which have been used to refer to them.
For the convenience of the reader, the various names and notations for these modules are summarized in the table \ref{tab-term} corresponding to the various objects acting on them.
\begin{table}[ht]
\caption{}
\begin{center}
\begin{tabular}{|c|c|c|c|}
 \hline
Representation & $\cal M$ & ${\cal M}_-$ & ${\cal M}_+$ \\ 
\hline\hline
$A_2$ & Fock space & odd & even \\
\hline
$Mp(4)$ & Metaplectic module & odd & even \\
\hline\hline
$OSp\,(1|4)$ & Singleton supermultiplet   
& Spinor singleton & Scalar singleton\\
\hline
$Spin(2,3)\cong Sp(4)$ &  Di\,-Rac module & Di :\quad ${\cal D}(1,\frac12)$ & Rac :\quad ${\cal D}(\frac12,0)$ \\
\hline
$Spin(1,3)\cong SL(2,{\mathbb C})$ &  Majorana representation  & Principal : $[\frac12,0]$ & Complementary : $[0,\frac12]$ \\
\hline
\end{tabular}
\end{center}
\label{tab-term}
\end{table}

It is also possible to reduce the  
complex
$\mathfrak{so}(2,3)$ algebra under its Lorentz subalgebra $\mathfrak{so}(1,3)\cong \mathfrak{sl}(2,\mathbb{R})\oplus\mathfrak{sl}(2,\mathbb{R})$ generated by

\begin{eqnarray}\label{SL2r-L}
J_0=\frac{\Gamma_0+S_{12}}{2},\qquad J_1=\frac{\Gamma_1+S_{02}}{2},\qquad J_2=-\frac{S_{01}-\Gamma_2}{2}\\[12pt] 
\bar{J}_0=\frac{\Gamma_0-S_{12}}{2},\qquad \bar{J}_1=\frac{\Gamma_1-S_{02}}{2},\qquad \bar{J}_2=-\frac{S_{01}+\Gamma_2}{2}.\label{SL2r-R}
\end{eqnarray}
Then, the module $\cal M$ decomposes into the tensor product of two $\mathfrak{sl}(2,\mathbb R)$ modules, say as ${\cal M}={\cal M}_L\otimes{\cal M}_R\,$.
The usefulness of the representations (\ref{SL2r-L})-(\ref{SL2r-R}) is that these submodules can be identified with the left and right subsectors, as is manifest when one makes use of the chiral representation of the Dirac matrices in (\ref{so32g}) or when one works with ``dotted'' and ``undotted'' spinors. In fact, the chiral representation is the one usually employed in the description of massless higher spin field on $AdS_4$ (\textit{e.g.} in \cite{Vasiliev:1995dn}) or on Minkowski space-time ${\mathbb R}^{1,3}$ (for instance in \cite{Klishevich:2001gy}).
It should be stressed however that $\mathfrak{so}(1,3)\cong \mathfrak{sl}(2,\mathbb R) \oplus \mathfrak{sl}(2,\mathbb R)$ is an isomorphism of \textit{complex} algebras but not of \textit{real} algebras. Roughly speaking, it does not preserve hermiticity nor the number of compact directions since the isomorphism makes use of multiplication by imaginary factors. If one complexifies $\mathfrak{so}(1,3)$ and defines $R_1=iJ_1 + i \bar J_1,
R_2=iJ_2 + i \bar J_2, R_3=J_0 + \bar J_0$ and
 $B_1=-i(iJ_1 - i \bar J_1),
B_2=-i(iJ_2 - i \bar J_2), B_3=-i(J_0 - \bar J_0)$, they generate the
Lorentz algebra, the $R$'s generating the rotations and the $B$'s
the boosts. But this representation is not unitary since these operators
are not Hermitian. More precisely, if one introduces 
$L_\alpha, \bar L_{\dot \alpha}$,
$\alpha,\dot \alpha=1,2$
such that $L_\alpha^\dag = \bar L_{\dot \alpha}$ and
$[L_1,L_2]=-i,[\bar L_{\dot 1},\bar L_{\dot 2}]=-i$ and define
$L_{\alpha \beta}= \frac12 \left\{L_\alpha,L_\beta\right\},
 \bar L_{\dot \alpha \dot \beta}= \frac12 \left\{\bar L_{\dot \alpha},
\bar L_{\dot \beta}\right\}$ and
$L_{\alpha  \dot \alpha}=\frac12 \left\{L_\alpha,\bar L_{\dot \alpha}\right\}$,
 the relationship with  $\mathfrak{so}(2,3)$
is given by $S_{\mu \nu} = \frac{i}{2} \sigma^{\mu \nu}_{\alpha \beta}
L^{\alpha \beta} - \frac{i}{2} \bar \sigma^{\mu \nu}_{\dot \alpha \dot \beta}
\bar L^{\dot \alpha \dot \beta}, \Gamma_\mu= 
\frac12 \sigma_\mu{}_{\alpha \dot \alpha} L^{\alpha \dot \alpha}$ with
$\sigma_\mu, \bar \sigma_\mu$ 
the Dirac matrices in the Weyl representation (see \cite{Klishevich:2001gy} 
for the notations). But since the $\sigma-$matrices are complex matrices
this clearly shows that the correspondance from $L_{\alpha \beta},
\bar L_{\dot \alpha \dot \beta}$ 
 and  $L_{\alpha \dot \alpha}$ to the $\mathfrak{so}(2,3)$ generators
only holds in a complexification.

The irreducible representations of the complex Lorentz algebra $\mathfrak{so}(3,1)$ are labelled by the spin of every $\mathfrak{sl}(2,\mathbb{R})$, \textit{i.e.} here the eigenvalues of $J_0=N_1/2$ and $\bar{J}_0=N_2/2$. 
This reproduces the same modules than for the reduction under the subalgebra $\mathfrak{so}(2)\oplus \mathfrak{so}(3)$, since their Cartan subalgebras are basically the same, 
formed by the number operators $N_1$ and $N_2$.
In Vasiliev's theory of interacting massless higher-spin fields on $AdS_4$ (as reviewed in \cite{Vasiliev:1995dn}), the higher-spin superalgebra of symmetries is isomorphic to the Weyl algebra $A_2$ (endowed with the commutator as Lie bracket). The metaplectic module is also a module of the Weyl algebra, thus the singleton supermultiplet is the module of the higher-spin superalgebra. The higher-spin superalgebra is then understood as the infinite-dimensional extension of the superalgebra $OSp(1|4)$ on $AdS_4$; it is realized as polynomials in the generators $L_a$ which are not constrained to be at most quadratic.
The elements of $A_2$ with degree higher than two are associated with the massless higher-spin fields on $AdS_4$. A somewhat surprising feature of the \textit{massive} higher-spin supermultiplet in Minkowski introduced in Section \ref{harmonic} later on is that it is a module of the higher-spin superalgebra on anti de Sitter. Though this group of symmetries acts only on the spinning degrees of freedom for the massive supermultiplet, it might be a remnant of the higher-spin gauge symmetries after a spontaneous symmetry breaking and a flat space-time limit. 

\section{Spectrum of the Majorana equation}\label{Majsp}

The Majorana theory \cite{Majorana} consists of the Dirac-like equation (\ref{Diraceq}) provided with the Poincar\'e generators (\ref{poincare}) and the spinning degrees of freedom given by the  $\mathfrak{so}(2,3)$ algebra representation (\ref{so32g}). 
Fermions and bosons are 
respectively contained in the Di and Rac modules of $\mathfrak{so}(2,3)$ and both can be unified in the supersymmetric singleton module of $\mathfrak{osp}(1|4)$. More concretely, the fields $\psi(x,q)$ are functions of the space-time coordinates $x^\mu$ and the internal coordinates $q_i$, the last providing the spinning degrees of freedom, completing effectively a six-dimensional configuration space.

The information about the particle content is encoded in the Casimir 
operators $P^2$ and the square of the Pauli-Lubanski vector (\ref{PLub}). 
For a single massive particle of mass $m$ and spin $J\,$, the quadratic and quartic Casimir operator of the Poincar\'e algebra $\mathfrak{iso}(1,3)$ are fixed and given by
$${\cal C}_2\Big(\mathfrak{iso}(1,3)\Big)=-P^2=m^2$$
and
\begin{equation}
{\cal C}_4\Big(\mathfrak{iso}(1,3)\Big)={\cal C}_2\Big(\mathfrak{iso}(1,3)\Big)\times{\cal C}_2\Big(\mathfrak{so}(3)\Big)
\quad\Longleftrightarrow\quad W^2=\,m^2\,J(J+1)\,.
\label{qrticC}
\end{equation}
Employing the identities (\ref{id}), one can show that, on the Di-Rac-Majorana representation, the squared Pauli-Lubanski vector takes the form
\begin{equation}
W^2=\frac{1}{4}\,P^2+(P\cdot\Gamma)^2.\label{Wsq1}
\end{equation}
Hence, in Majorana's theory a particle with some fixed spin and mass is equivalently characterized by the equations
\begin{equation}\label{KGMaj}
(P\cdot \Gamma-M)\psi(x,q)=0, \qquad (P^2+m^2)\psi(x,q)=0.
\end{equation}
 The first one corresponds to the infinite-component Majorana equation and the second one is the Klein-Gordon equation. In Majorana's construction, these equations are completely independent, as explained in Section \ref{anydim}.
According to the values of $P^2$, massless, tachyon and massive particles can be 
obtained.\footnote{See ref. \cite{Sudarshan:1970ss}, and appendix \ref{app} for a summarized account of the massless and tachyonic solutions.}  The spectrum of particles in every sector (summarized in table \ref{tab-maj}) can be obtained from the representation of the  little algebra in the standard frame, 
which is contained in the respective stabilizer of  $\mathfrak{so}(2,3)$ (see table \ref{tab-ads} in Appendix \ref{app}).

\begin{table}[ht]
\caption{}
\begin{center}
\begin{tabular}{|c|c|c|c|c|}
 \hline
            UIR   &  Standard frame & Stability & Majorana  \\
                  &    $P^\mu$       & algebra   & spectrum   \\\hline\hline
\small{Massive}   & $(m,0,0,0)$  & $\mathfrak{so}(3)$   & $m=M/(J+\frac{1}{2})$ : $J=0,1/2,1,...$\\\hline
\small{Continuous-spin}  & $(E,0,0,E)$  & $\mathfrak{iso}(2)$  &  $E=M/\varepsilon$ : ${ 0} <\varepsilon < + \infty$\\\hline
\small{Tachyonic} & $(0,0,0,\ell)$  & $\mathfrak{so}(2,1)$ & 
$\ell=M/\sigma$ : ${ 0<|\sigma|} <\infty$\\\hline
\end{tabular}
\end{center}
\label{tab-maj}
\end{table}
For our purpose, we focus ourself here in the study of the massive and massless solutions.

In the rest frame for a massive particle of mass say $m\,$, the momentum is $P^\mu=(m,0,0,0)$ and the Dirac-like operator reads $P\cdot\Gamma = m\,\Gamma_0\,$. The spectrum of eigenvalues of the operator $\Gamma_0$ is the set of non-vanishing half-odd-integer numbers, so the Majorana equation produces the mass spectrum
\begin{equation}
m_J=\frac{M}{J+\frac{1}{2}},\qquad J=0,1/2,1, \cdots \label{MassSpectr}
\end{equation}
Observe that there are only positive-energy solutions. 
Moreover, for any fixed momenta the eigenpaces of $\Gamma_0$ is isomorphic 
to the $\mathfrak{so}(3)$-module ${\cal D}_J$ of dimension $2J+1$ 
(see Appendix \ref{app}, \eqref{gamma0}). Therefore, the half-integer $J$ gives the spin of the corresponding particle of mass $m_J\,$ and the operator ${\cal R}$ (\ref{RR'}) is the statistical phase (see (\ref{phase})).

The statistical phase for every particle is also given by
(\ref{phase}) with $J$, the eigenvalues of $\hat{S}$.
A specific particle in this spectrum can be extracted out by imposing simultaneously  the Dirac-like equation in (\ref{KGMaj}) and the Klein-Gordon equation $P^2+m_J^2=0\,$. Differential equations of different sort can be proposed such that a finite 
set of particles in the massive sector of the Majorana representation are 
extracted.
 On the one hand, first order field equations have already been given for spin zero 
in \cite{Dirac} (the so-called ``new Dirac equation'') and spin $1/2$ in \cite{Stau2} 
(the ``Staunton equation''). On the other hand, Bidenharn proposed \cite{BiedHan} a differential equation of order $2J+1$ which extracts all the massive particles with spin $\leq 2J$. All these equations have only positive energy solutions.

 In the standard frame for a massless particle of energy $E\,$, the momentum 
is $P^\mu=(E,0,0,E)$ and the Dirac-like operator reads $P\cdot\Gamma = E\,(\Gamma_0+\Gamma_3)\,$. The spectrum of eigenvalues of the operator 
$\Gamma_0 + \Gamma_3$ is the subset of real numbers $\varepsilon>0$. 
In order to understand the massless spectrum, it is better to look at the squared Pauli-Lubanski (\ref{Wsq1}) which in the present case is equal to
$W^2=(P\cdot\Gamma)^2=E^2\varepsilon^2 =M^2$ by virtue of the Majorana equation. This means that the massless sector is made of the ``continuous-spin'' representations parametrized by $M$. A more detailed analysis (see \cite{Sudarshan:1970ss} and Appendix \ref{app}) shows that the massless spectrum is actually made of the direct sum of the bosonic and the fermionic continuous-spin representations.
Khan and Ramond provided an enlightening interpretation of the elusive ``continuous-spin'' (bosonic or fermionic) representations 
as the limit of massive representations where the (either integer or half-odd-integer) spin goes to infinity, $J\rightarrow\infty$ and 
the mass to zero, $m\rightarrow 0$, with their product $J\,m=M$ kept fixed to a constant $M$ with the dimension of a mass \cite{Bekaert:2005in}. Indeed, one can check that
the squared Pauli-Lubanski has the limit: $W^2=m^2J(J+1)\rightarrow M^2\,$.
Remark that the Majorana equation provides a particular realization of this limit because the massless sector can be understood as the infinite-spin limit of the massive sector, see (\ref{MassSpectr}).
More generally, this new interpretation explains many exotic features of the continuous-spin representations. For instance, although they are massless the continuous-spin representations are not conformally invariant since they are characterized by a parameter with the dimension of a mass, like massive particles.

It is worth mentioning that, formally, the (spin $1/2$) Dirac equation 
consists in replacing the $\Gamma_\mu$ operator by the Dirac-matrices ($-\small{\frac{i}{2}} \gamma_\mu$ more precisely), and the continuous internal variables $q_i$ by discrete (spinor) variables. Other finite representations for $\Gamma_\mu$ can be considered, producing other equations, like the Kemmer-Duffin-Petiau (describing a scalar and a vector massive field) for example. In the general case, studied by Bhabha \cite{Bhabha:1945zz}, it is not possible to fix
univocally the Poincar\'e representation, and a finite number of particles appear. A characteristic of these systems is the increasing number of particles  and the maximal spin in the spectrum, when the dimension of the $\Gamma$-matrices increases, while the mass still decreases with the increasing spin. It is interesting to observe that the Dirac and the Majorana equations look like limiting cases, \textit{i.e.} when the dimension of the $\Gamma$-matrices is four (the minimum possible to get a faithful $\mathfrak{so}(3,2)$ representation) and, respectively, when their dimension is infinite. They correspond also to two ``extreme'' cases when the equation produces either only one irreducible representation of the Poincar\'e group, describing a fermion, 
or when the representation contains an infinite number of particles.

\section{Relativistic wave equations and arbitrary Regge trajectories}\label{Reggelike}

The Majorana UIRs of the   Lorentz algebra can be used to describe an arbitrary number of
particles with spin and mass distributed in arbitrary (non-degenerate) Regge 
trajectories. Group theoretically, that can be done by imposing some relation between the Casimir operators of the Poincar\'e
algebra. Field theoretically, this can be realized by establishing a functional relation between the Dirac-like operator $P\cdot\Gamma$ and the momentum square $P^2\,$. Then, the Poincar\'e group
representation is not irreducible but decomposes as direct sum of UIRs with a certain
correlation between spin and mass. The constraint has the general form,
\begin{equation}
\label{Regge0}
{\cal F } (P\cdot\Gamma,P^2)\psi(x,q)=0,
\end{equation}
such that the infinite-component Majorana equation and the Klein-Gordon equation can respectively be seen as the two particular cases where the function ${\cal F }(y,z)$ is either linear in $y$ or $z$ either leading to a decreasing Regge trajectory (\ref{MassSpectr}) or to a horizontal Regge trajectory $m^2$ = constant. The first example corresponds to Majorana's theory while the second corresponds to the massive higher-spin supermultiplet, as discussed in many detail in the next section. 
A general discussion of wave equations with the form ${\cal F } (P\cdot\Gamma,P^2)=f(P^2)-(P\cdot\Gamma)^2$ was provided in \cite{Casalbuoni:1971nj}. We will not attempt to present a completely exhaustive discussion of infinite-component wave equations (\ref{Regge0}) based on the Majorana representation but we will merely describe a particular prescription for describing Regge trajectories.
In order to ensure a spectrum of massive particles, one can assume for instance
\begin{equation}
\label{Regge1}
{\cal F } (P\cdot\Gamma,P^2)=P^2+{\cal G}_+(P\cdot\Gamma,P^2),
\end{equation}
where  ${\cal G}_+>0$ is a positive definite function on $\mathbb{R}^2$, such that the solutions of (\ref{Regge0})  can be only massive $P^2<0$.
Notice that, when restricted to massive solutions $P^2<0$, the square of the Pauli-Lubanski vector (\ref{Wsq1}) can be written as,
\begin{eqnarray}\label{Wsq2}
W^2=-P^2\hat{J}(\hat{J}+1)\,,\qquad \hat{J}:=
\frac{P\cdot\Gamma}{\sqrt{-P^2}}-1/2\,,
\end{eqnarray}
from where we identified the ``spin operator'' $\hat{J}$ (observe that this operator is non-local) by analogy with equation (\ref{qrticC}). Passing to the standard frame $P^\mu=(m,0,0,0)$ this operator becomes equivalent to the Hamiltonian of a harmonic oscillator on the plane 
(see (\ref{sp4}) and (\ref{u1xsu2})),
\begin{equation}
\hat{J}\stackrel{\hbox{\tiny{rest}}}{=}\hat{S}=\frac{\hat{H}-1}{2} \qquad \Rightarrow \qquad \hat{J}=0,1/2,1,\cdots \label{J=S}
\end{equation}
hence, it takes only half-integer eigenvalues.
We can rewrite now (\ref{Regge1}) in terms of the spin operator via
$${\cal G}_+\Big(P\cdot\Gamma,P^2\Big)={\cal G}_+\Big(\sqrt{-P^2}(\hat{J}+1/2),P^2\Big)\,,$$
where the apparent non-locality is just an artifact coming from the use of the spin operator.
In order to produce arbitrary non-degenerate Regge trajectories, one should assume that the corresponding equation $${\cal F}\Big(m(J+1/2),m^2\Big)=0$$ can, say, be solved for the spin $J$ in terms of the mass $m$ (this requires some invertibility condition such as $\partial{\cal G}_+(y,z)/\partial y\neq 0$). These rising Regge trajectories requires either a non-local or at least a higher-derivative wave operator ${\cal F } (P\cdot\Gamma,P^2)$ whose corresponding inverse (\textit{i.e.} the propagator) would give rise to unphysical properties such as ghosts, unfortunately. This is in agreement with the no-go theorem \cite{Grodsky:1968zz}.
In particular, the wave equation where $\alpha^\prime$ with dimension of length square and $c>0$ a dimensionless positive constant (\textit{cf.} \cite{vanDam:1978xd})
$$
\alpha^\prime\,P^2\,+\,c+\hat{J}=0,
$$
produces a linearly rising Regge trajectory $m^2=(J+c)/\alpha^\prime$
(``Chew-Frautchi plot'') but is highly non-local. Nevertheless, the local (but sixth order) equation
$$P^2\Big(\,\alpha^\prime P^2\,+\,c-\frac{\,1}{\,2}\,\Big)^2+\big(P\cdot\Gamma\big)^2=0,$$
reproduces the same linearly rising Regge trajectory.

\section{Massive higher spin fields as covariant harmonic oscillator}\label{harmonic}


Let us turn back to the particular case of a wave function taking values in Majorana's representation of the Lorentz group and satisfying  the Klein-Gordon equation. The corresponding mass spectrum has infinite degeneracy and corresponds to a horizontal Regge trajectory. In this simplest case, there exists a simple dictionary between the Fock space of a planar harmonic oscillator and the Hilbert space of positive-energy particles in this horizontal Regge trajectory.

The spin operator in the rest frame \eqref{J=S} takes a particularly 
simple expression, corresponding to the Hamiltonian of a planar harmonic oscillator.
The raising operators $a_i^+$ transform as two-component Weyl spinors under the rotation subalgebra $\mathfrak{so}(3)$ so that it is clear that the basis elements $a_{i_1}^+\ldots a_{i_{2J}}^+\mid\,0\rangle$ of the Fock space at level $J$ span the spin-$J$ module ${\cal D}_J$ of $\mathfrak{so}(3)\,$.
Thus to any energy level with
degeneracy equal to $2 J+1=N_1+N_2+1$ ($N_1,N_2\in\mathbb N$) in the Fock space corresponds a 
massive particle of spin $J=(N_1+N_2)/2$ in the Hilbert space. These degenerated states, corresponding to $(a_1^+)^{N_1}({a_2}^+)^{N_2}\mid\,0\rangle$
in the Fock space, can be labelled by the third component of the spin, say $S_3=S_{12}=(N_1-N_2)/2$.
The raising $a_i^+$ and lowering $a_i^-$ operators ($i=1,2$) respectively increase and 
decrease the energy of the planar harmonic oscillator by one unit.
Since the spin is related to the energy in
\eqref{gamma0}, this means that
the same operators acting on the Hilbert space equivalently increase or decrease the spin by\footnote{More precisely, the oscillators $a_1^\pm$ create/destroy a spin ``up'' while the oscillators $a_2^\pm$ create/destroy a spin ``down.''} one-half, 
i.e. they correspond to a supersymmetry
transformation. As mentioned in the Appendix \ref{app} this ``bosonic''
realisation of supersymmetry which comes from the Weyl algebra $A_2$ has
the interesting consequence that the irreducible supermultiplets may contain an 
infinite number of particles. This  motivates to establish the precise
dictionary between the harmonic oscillator on the plane 
and the massive higher-spin supermultiplet. 
We would like to have therefore, the Poincar\'e covariant equivalent 
expression
of the harmonic oscillators  as well as the eigenstates of the Hamiltonian.
For that purpose, 
we define the covariant version of the  creation and annihilation
 operators (in the sense that they reduce 
 in the rest frame to $a_i^\pm$) as follows:

\begin{equation}\label{D+-}
D^{\pm}_a=(\pm iP^\mu\gamma_\mu+\sqrt{-P^2})_a\,^bL_b.
\end{equation}
They satisfy  $(D^-_a)^\dagger=D^+_a$.
In the rest frame $P^\mu=(m,0,0,0)$ we have

\begin{equation}\label{D+-rest}
[D^{\pm}_a]_{rest}=m(a_1^\pm, a_2^\pm ,\pm i a_1^\pm ,
\pm i a_2^\pm).
\end{equation}

\noi
A simple computation gives 
\begin{eqnarray}
\label{DD}
&[D^{\pm}_a,D^{\pm}_b]=0,\qquad [D^+_a,D^-_b]=2
\sqrt{-P^2}(P\cdot\gamma)_{ab}-2iP^2C_{ab},&\label{com:D-D}\\[10pt]
&[{ J}_{\mu\nu},D^{\pm}_a] =-(\gamma_{\mu\nu})_a\,^b D^{\pm}_b,
\qquad [P_\mu,D^{\pm}_a]=0.&
\end{eqnarray} 
This algebra by itself does not require the mass-shell condition.
Observe however that ``passing to the rest frame,'' states with different masses would produce the same vacuum for the harmonic oscillator, by means of the equation $a_i^-|\,0\rangle=0$. In other words, the ``covariant'' vacuum defined by $D^-_a\phi(x,q)=0$, has an indeterminate mass. The mass-shell 
condition is a first class constraint consistent with 
covariant-vacuum equation since $[P^2+m^2,D^-_a]=0$.
Thus it can be used to redefine 
$\tilde{D}^{-}_a:=[D^{-}_a]_{\sqrt{-P^2}=m}=(-iP^\mu\gamma_\mu+m)_a\,^bL_b$. In this way, the equation
\begin{equation}\label{Dirac}
\tilde{D}^{-}_a\phi(x,q)=0,
\end{equation}
produces a well defined {\it covariant vacuum}  with spin zero 
and mass $m$, which in fact turns to be
the  ``new Dirac equation'' \cite{Dirac}. We can verify the consistency 
of the  condition (\ref{Dirac}). Indeed the commutator 
$[\tilde{D}^{-}_a,\tilde{D}^{-}_b]={  -i}C_{ab}(P^2+m^2)$ together with 
 $L^a\tilde{D}^{-}_a=-4i(P\cdot\Gamma-m/2)$
leads to the Klein-Gordon and Majorana equations 
\begin{equation}\label{phi}
(P^2+m^2)\phi(x,q)=0,\qquad (P\cdot\Gamma-m/2)\phi(x,q)=0.
\end{equation}
By (\ref{Wsq2}), $\phi(x,q)$ has mass $m$ and spin $0$. The 
vacuum has been  defined, so the raising
operator $D^+_a$ enables us to get higher-spin particles. In fact, the fields
\begin{equation}\label{S-field}
\varphi_S (x,q)=\zeta^{a_1a_2\dots a_{2S}}
\varphi_{a_1a_2\dots_{2S}}(x,q),\qquad 
\varphi_{a_1a_2\dots
a_{2S}}(x,q)=D^{+}_{a_1}D^{+}_{a_{2}}\dots D^{+}_{a_{2S}}\phi(x,q),
\end{equation}
where $\zeta^{a_1a_2\dots a_{2S}}$ are  constant symmetric spinor-tensors, 
have mass $m$ and spin $S$, where $S=0,1/2,1,\cdots$,
since they satisfy the Klein-Gordon and the {\it spin equation},
\begin{eqnarray}\label{masespineq}
(P^2+m^2)\varphi_S (x,q)=0,\qquad (\hat{J}-S)\varphi_S (x,q)=0.
\end{eqnarray}
Taking in account (\ref{phi}) for the vacuum, the first relation is automatic from
$[P^2+m^2,D^{+}_{a_1}D^{+}_{a_{2}}\dots D^{+}_{a_{2S}}]=0$ whereas
the second one is obtained considering
\begin{equation}\label{JD}
[\hat{J},D^{\pm}_{a}]=\pm 1/2 D^{\pm}_{a}, 
\end{equation}
and $[\hat{J},D^{+}_{a_1}D^{+}_{a_{2S}}\dots D^{+}_{a_{2S}}]=S
D^{+}_{a_1}D^{+}_{a_{2}}\dots D^{+}_{a_{2S}}$. 
The field (\ref{S-field}) is completely symmetric in the $a_i$ indices, since $D^{+}_{a_1}$ are commuting operators. Furthermore,
observe that supersymmetric transformations are generated by the $D^{+}_a$ and
$D^{-}_a$ operators, as they increase or decrease the spin in
one-half unit

$$
(\hat{J}-(S\pm1/2))D^{\pm}_a\varphi_S (x,q)=0,
$$

\noi
as a simple consequence of (\ref{JD}) and the fact that
 $\varphi_S (x,q)$ is a spin-$S$ field, i.e. a solution of (\ref{masespineq}). 
The general solution of the Klein-Gordon equation is therefore 
\begin{equation}\label{s-multpt}
\varphi(x,q)=\sum \limits_{S\in\{0,1/2,1,\cdots\}}
 \varphi_S (x,q), 
\end{equation}
with every $\varphi_S (x,q)$, carrying an
irreducible representation of the Poincar\'e group of spin $S$. 
This   infinite tower of higher spin
is indeed a consequence of the decomposition 
$\mathfrak{so}(2) \oplus \mathfrak{so}(3) 
\subset \mathfrak{so}(3,2)$ \eqref{massive}, 
and  turns out to be a  characteristic of the 
Majorana representation.
These fields are solutions of the Dirac-Fierz-Pauli \cite{Fierz} equations for massive fields. Let us define  the operators,

\begin{eqnarray}
&\hat{\phi}^\pm_{\mu_1\mu_2 ...\mu_n} =  \hat{h}^\pm_{\mu_1} ...\hat{h}^\pm_{\mu_n}, \qquad \hat{\Psi}^\pm_{\mu_1\mu_2 ...\mu_n}{}^a = \hat{h}^\pm_{\mu_1} ...\hat{h}^\pm_{\mu_n}D^\pm{}^a,& \label{hatfield}\\[12pt]
&\hat{h}^\pm_\mu= (\gamma_\mu)^{ab}D^\pm_aD^\pm_b= 8 \left(-P^2 \Gamma_\mu + 
P\cdot\Gamma P_\mu \pm i \sqrt{-P^2}S_{\mu\nu}P^\nu\right).&\label{h+-}
\end{eqnarray}

\noi
From (\ref{S-field}), saturating the spinor indices by contraction with the
$\gamma_\mu$-matrices one gets the fields,
\begin{eqnarray}
\phi_{\mu_1\mu_2 ...\mu_n}(x,q):=\hat{\phi}^+_{\mu_1\mu_2
...\mu_n}\phi(x,q), \qquad
\Psi_{\mu_1\mu_2 ...\mu_n}{}^a(x,q):= \hat{\Psi}^+_{\mu_1\mu_2 ...\mu_n}{}^a
\phi(x,q). \label{fields}
\end{eqnarray}
By construction, they have integer spin-$S=n$ or half-odd-integer spin-$S=n+1/2$ respectively, since
{
\begin{equation}\label{Jhatfield}
[\hat{J},\hat{\phi}^\pm_{\mu_1\mu_2
...\mu_n}]=\pm n\hat{\phi}^\pm_{\mu_1\mu_2
...\mu_n} , \qquad [\hat{J},\hat{\Psi}^\pm_{\mu_1\mu_2 ...\mu_n}{}^a]=\pm(n+1/2)\hat{\Psi}^\pm_{\mu_1\mu_2 ...\mu_n}{}^a.
\end{equation}
} Now, from the identities,
\begin{eqnarray}
&(-iP^\mu\gamma_\mu+m)_a\,^c (+iP^\mu\gamma_\mu+m)_c\,^b=(P^2+m^2)
\delta_a^b,&\label{id-fer}\\[12pt]
& iP^\mu \hat{h}^\pm_\mu=\partial^\mu \hat{h}^\pm_\mu=0,\qquad \hat{h}^\pm {^\mu}
\hat{h}^+_\mu=0,&\label{id-bos}
\end{eqnarray}
we check easily that fields (\ref{fields})
are also solutions of the higher spin field equations\footnote{In order to prove the last equation in \eqref{DFPfer}, it is necessary to use the fact that the operator $(\gamma^\mu)_a{}^b\hat{h}^+_\mu D^+{}_b$ vanishes. This fact can be checked in the rest.} 
\begin{eqnarray}
&(P^2+m^2)\phi_{\mu_1\mu_2 ...\mu_n}(x,q)=0, \qquad
 \partial^\mu\phi_{\mu\mu_1 ...\mu_{n-1}}(x,q)=0,\qquad \phi^\mu{}_{\mu\mu_2
 ...\mu_{n-2}}(x,q)=0,& \label{DFPbos}\\[12pt]
&(iP^\mu\gamma_\mu-m)_a\,^b(\Psi_{\mu_1 ...\mu_n})_b(x,q)=0,\,
\Psi^\mu{}_{\mu \mu_1 ...\mu_{n-2}}(x,q)=0, \,
(\gamma^\mu)_a\,^b(\Psi_{\mu \mu_1 ...\mu_{n-1}})_b(x,q)=0,& \label{DFPfer}
\end{eqnarray}
but they have only positive energy, by construction. Observe however that a dependence in the internal space $\mathbb{R}^2 \ni (q_1,q_2)$ still remains.

\noi
In summary, via ``covariantization'' we have established the precise
dictionary between the states of the planar harmonic oscillator
and the positive-energy  massive particles in a horizontal Regge trajectory including all spins.

\begin{table}[ht]
\caption{}
\begin{center}
\begin{tabular}{|c|c|c|c|c|c|}
\hline
Planar harmonic oscillator   &  Horizontal Regge trajectory  \\\hline
$a_i^\pm$  & $D_a^\pm$ \\
$a_i^-|\,0\rangle=0$ & $D_a^-\phi(x,q)=0, \quad (P^2+m^2)\phi(x,q)=0$\\
$|n_1,n_2\rangle$ & $\varphi_S(x,q),\quad S=\frac{n_1+n_2}{2}$\\
$H$               & $\hat{J}+1/2$\\
$\mathfrak{so}(3)$ & $\mathfrak{so}(1,3)$\\
\hline
\end{tabular}
\end{center}
\end{table}

\section{Symmetries of the infinite set of fields} \label{S-P sec}

We have already seen that the multiplet (\ref{s-multpt}) exhibit some aspect 
of supersymmetry since it contains
 bosons and fermions which are related by the $D_a^\pm$ 
operators. Because it contains an infinite number of particles, 
in principle they could involve additional ``internal'' 
symmetries. We can immediately see for instance, that the Hermitian 
combination of operators $D^\pm_a$ gives rise to
$$
L_a=\frac{D^+_a+D^-_a}{\sqrt{-P^2}},
$$
i.e., the $\mathfrak{osp}(1|4)$ supercharge. Hence, the
supermultiplet (\ref{s-multpt}) carries not only representation of
the Poincar\'e algebra, but also a representation of 
$\mathfrak{osp}(1|4)$. This supermultiplet can be considered as the positive-energy sector of
of the equation (\ref{Regge1}) in the particular case when ${\cal G}_+=m^2$ (\textit{i.e.} a ``horizontal'' Regge-trajectory) or like the the limit $J\rightarrow \infty$ of the Biedenharn equation \cite{BiedHan}. In fact the Biedenharn equation, is a generalization of the New Dirac equation \eqref{Dirac}, which can be written as
$$
\tilde{D}^-_{a_1}...\tilde{D}^-_{a_{2J+1}}\phi(x,q)=0,\qquad J=0,1/2,1,3/2,..., 
$$
whose solutions are all fields with spin $S\leq J$ and mass $m$.
At first glance however, the operators $D^{\pm}_a$
do not satisfy the usual anticommutation relation of supersymmetry

\begin{eqnarray}
&\{D^\pm_a,D^\pm_b\}=4 \sqrt{-P^2}
\hat{A}^\pm_\mu(\gamma^\mu)_{ab}\pm 4i
\hat{F}^\pm_{\mu\nu}(\gamma^{\mu\nu})_{ab},&\label{anticDD}\\[12pt]
&\hat{A}^\pm_\mu :=\sqrt{-P^2} \Gamma_\mu +
\frac{P\cdot\Gamma}{\sqrt{-P^2}}\,P_\mu \pm i S_{\mu\nu}P^\nu, \qquad
\hat{F}^\pm_{\mu\nu}:=\partial_\mu \hat{A}^\pm_\nu - \partial_\nu
\hat{A}^\pm_\mu.&\label{A-F}
\end{eqnarray}
The equation (\ref{anticDD}) should be compared with (\ref{h+-}) where $\hat{h}_\mu^\pm=8\sqrt{-P^2}\hat{A}_\mu^\pm$. The vector and antisymmetric tensor operators $\hat{A}^\pm_\mu$ and $\hat{F}^\pm_{\mu\nu}$ are related as a vector field and its fieldstrength. Actually the analogy may be pursued further because they obey to the Proca-like identity: $\partial^\mu \hat{F}_{\mu\nu}-\partial^\mu \partial_\mu \hat A_\nu\equiv0$. Observe that here we use a Schr\"odinger representation of these operators (acting on the internal space), which seems to be quite analogous to the representation of the three-dimensional Abelian Chern-Simons field in \cite{Dunne:1989cz}. 
It might be useful to provide several interpretations of the algebra (\ref{anticDD}). 

Firstly, operators in the r.h.s of (\ref{anticDD}) carry spin one,
$$
[\hat{J},\{D^\pm_a,D^\pm_b\}]=\pm \{D^\pm_a,D^\pm_b\}.
$$
That means, both $\hat{A}^\pm$ and $\hat{F}^\pm$ acting on a field of 
spin $S$ field produce another one of spin $S\pm1$. 
In particular, acting on the vacuum, they produce 
\begin{equation}
 A_\mu(x,q):= \hat{A}^+_\mu \phi(x,q),\qquad F_{\mu\nu}(x,q):= \hat{F}^\pm_{\mu\nu}\phi(x,q)=\partial_\mu A_\nu - \partial_\nu
A_\mu. 
\end{equation}
Then, owing the identities (\ref{id-bos}) one can check that the Proca equation and the  transversality condition are satisfied
\begin{equation}\label{proca}
\partial^\mu F_{\mu\nu}-m^2 A_\nu=0, \qquad \partial^\nu A_\nu=0.
\end{equation}

\noi
Of course, this interpretation applies also for all operators $\hat{\phi}^+$ and $\hat{\Psi}^+$ in (\ref{hatfield}). 
Thus, $D_a^\pm$ are thus a kind of square root of the
massive vector fields $A_\mu^\pm$.

Secondly, let us rewrite the anticommutator (\ref{anticDD}),
$$
\{D^\pm_a,D^\pm_b\}=
 \left(4\sqrt{-P^2}\hat{J} P_\mu\right)(\gamma^\mu)_{ab} \pm 4i
\hat{F}^\pm_{\mu\nu}(\gamma^{\mu\nu})_{ab}+\left(-4P^2 \Gamma_\mu +2\sqrt{-P^2}P_\mu  \pm i 4\sqrt{-P^2}S_{\mu\nu}P^\nu\right)(\gamma^\mu)_{ab}
$$
and consider the ``continuous-spin limit'' \cite{Bekaert:2005in}:
$\sqrt{-P^2}\rightarrow 0$, $\hat{J}\rightarrow \infty$ such that $\sqrt{-P^2}\hat{J}= M=constant$, then the last term in parenthesis vanishes and the anticommutator produces, 
\begin{equation}
\{Q^\pm_a,Q^\pm_b\}= 2 P_\mu(\gamma^\mu)_{ab} \pm 2i \hat{W}^\pm_{\mu\nu}(\gamma^{\mu\nu})_{ab}, \label{QQ}
\end{equation}
 which are the usual anticommutation relation of the 
superPoincar\'e algebra with tensorial central charges.
\noi 
Here we have defined 
\begin{equation}\label{QWdef}
Q^\pm_a:=(\sqrt{2M})^{-1}D^\pm_a \qquad \hbox{and} \qquad \hat{W}^\pm_{\mu\nu}:=(\sqrt{2M})^{-1}\hat{F}^\pm_{\mu\nu}.
\end{equation}
The operator {$$\hat{W}^\pm_{\mu\nu}=i(\sqrt{2M})^{-1}\Big( \sqrt{-P^2} (P_\mu \Gamma_\nu-P_\nu \Gamma_\mu)\pm i (P_\mu S_{\nu \lambda}-P_\nu S_{\mu \lambda})P^\lambda\Big)$$}(whose first term in parenthesis vanishes in the massless limit) is a central antisymmetric Lorentz tensor of rank two. Observe that this limit is equivalent to consider the large spin solutions of the Majorana equation. That comes from the condition $\sqrt{-P^2}\hat{J}=P\cdot\Gamma-\sqrt{-P^2}/2=M$, which in the continuous-spin limit is equivalent to $P\cdot\Gamma=M$, the Majorana equation. It shows that the continuous-spin limit of the massive solutions of the Majorana equation and its massless solutions are equivalent. Hence, the massless sector of the Majorana equation form a (tensorial central extended) superPoincar\'e multiplet (\textit{c.f.} \cite{Horvathy:2006pw,Brink:2002zx}). It can be seen also, simply by checking that the supercharges $Q^\pm_a$ are observables with respect to the Majorana equ
 ation for massless particles, \textit{i.e.}, the commutator
\begin{equation}\label{susybr}
[P\cdot\Gamma-M,Q^\pm_a]=\pm \sqrt{-P^2}Q^\pm_a /2 
\end{equation}
vanishes for massless particles.
Thirdly, we can change our point of view and
rewrite the anticommutator of $Q^\pm_a$  
$$
\{Q^\pm_a,Q^\pm_b\}= 2 {\cal P}_\mu^\pm(\gamma^\mu)_{ab}
 \pm 2i \hat{W}^\pm_{\mu\nu}(\gamma^{\mu\nu})_{ab},$$
where
\begin{equation}
\label{translgen}
{\cal P}_\mu^{ \pm} :=(\sqrt{2M})^{-1}\sqrt{-P^2}\hat{A}^+_\mu=(\sqrt{2M})^{-1} \left((P\cdot\Gamma) P_\mu-P^2 \Gamma_\mu \pm i \sqrt{-P^2} S_{\mu\nu}P^\nu \right). 
\end{equation}
If we interpret
either ${\cal P}_\mu^+$ or ${\cal P}_\mu^-$ as the ``translation'' operator, this is also the anticommutation relation of the superPoincar\'e algebra with tensorial central charges. In fact,
\begin{eqnarray}\label{salg}
&[{ J}_{\mu\nu},{
J}_{\lambda\rho}]=i(\eta_{\mu\lambda}{ J}_{\nu\rho} +
\eta_{\nu\rho}{ J}_{\mu\lambda}- \eta_{\mu\rho}{
J}_{\nu\lambda}- \eta_{\nu\lambda}{ J}_{\mu\rho}),& \nonumber\\[12pt]
&[{ J}_{\mu\nu},{\cal P}_\lambda^\pm]=
i(\eta_{\mu\lambda} {\cal P}_\nu^\pm-
\eta_{\nu\lambda}{\cal P}_\mu^\pm), 
\quad [{\cal P}_\mu^\pm,{\cal P}_\nu^\pm]=0,\nonumber&\\[12pt]
&[{ J}_{\mu\nu},Q^\pm_a]=-(\gamma_{\mu\nu})_a\,^bQ^\pm_b, \qquad
[{\cal P}_\mu^\pm,Q^\pm_a]=0.&\label{Palg}\\[12pt]
&[{ J}_{\mu\nu},\hat{W}^\pm_{\lambda\rho}]=i(\eta_{\mu\lambda}\hat{W}^\pm_{\nu\rho} +
\eta_{\nu\rho}\hat{W}^\pm_{\mu\lambda}- \eta_{\mu\rho}\hat{W}^\pm_{\nu\lambda}- \eta_{\nu\lambda}\hat{W}^\pm_{\mu\rho}),& \nonumber\\[12pt]
&[\hat{W}^\pm_{\mu\nu},\hat{W}^\pm_{\lambda\rho}]=0,\qquad [\hat{W}^\pm_{\mu\nu},{\cal P}_\lambda^\pm]=0,
\qquad [\hat{W}^\pm_{\mu\nu},Q^\pm_a]=0.&\nonumber
\end{eqnarray}

\noi
Some remarks are in order here. The generators
with the same upper index $\pm$ commute with each other, as a simple consequence of
\eqref{DD}. However, since $[D_a^+,D_b^-]
\ne 0$ the operators with different upper indices do not commute. Choosing just one type of operators in \eqref{salg}, \textit{i.e.} with the upper index $+$ or $-$, one could be tempted to interpret $\hat{W}^\pm_{\mu\nu}$ either, like the
electric-magnetic charge of a membrane or like
translation generators in a tensorial space. On the one hand, $\hat{W}^\pm_{\mu\nu}$, being related the 
fieldstrength of the Proca field (\ref{proca}), seems to be closer to the first interpretation. On the other hand, ${\cal P}_\mu^\pm$ and $\hat{W}^\pm_{\mu\nu}$ together, could be seen as the translation operators in a 
$4+6$ dimensional (tensorial) space.

We would like to study the algebraic structure of operators 
(\ref{hatfield}). Because of \eqref{DD}, they commute  when 
they have the same upper indices, symbolically; $[\hat{\phi}^\pm, \hat{\phi}^\pm]=[\hat{\phi}^\pm, \hat{\Psi}^\pm]=[\hat{\Psi}^\pm, \hat{\Psi}^\pm]=0$. But when they have different upper indices, as $Q_a^+$ and ${\cal P}_\mu^-$
{\it etc.}, they satisfy a complicated algebra. It is better therefore to proceed by looking at their algebraic properties, starting with the rough structure and then depurating  it.
\begin{enumerate}
\item Firstly, consider the infinite-dimensional associative algebra presented by the generators $D^+_a$ and $D^-_b$ modulo the (anti)commutation relations \eqref{DD} where $P^\mu$ are seen as numbers:
\begin{equation}
{\cal C}A_2(P):=\{\hbox{All possible products of } D^+_a , D^-_a
 \} \label{CA2} 
\end{equation}
In the particular case $P^\mu=(m,0,0,0)$, it becomes the Weyl algebra, so we could refer to ${\cal C}A_2(P)$ as the ``covariantized Weyl algebra''. 

\item Secondly, we can notice that ${\cal C}A_2(P)$ is ${\mathbb Z}_2$-graded and
$${\cal C}A_{2,0}(P):=\{\hbox{All even order products} \in {\cal C}A_2(P) \},$$
$${\cal C}A_{2,1}(P):=\{\hbox{All odd order products} \in {\cal C}A_2(P) \}.$$
Hence 
\begin{equation}
[{\cal R},{\cal C}A_{2,0}]=0,\qquad  \{{\cal R},{\cal C}A_{2,1}\}=0.
\end{equation}
The subspace ${\cal C}A_{2,0}(P)$ corresponds
 to the operators which do not change the statistics of the fields, since it commutes with the ``statistical phase'' ${\cal R}$. The operators of ${\cal C}A_{2,1}$ change the statistics, 
shifting the spin by a half-odd-integer number, acting as generalized supercharges.

\item Thirdly, we can extract the operators which change the spin in a fixed 
quantity
 \begin{eqnarray}
&\phi^\pm_{(n)}:=\Big\{\hbox{All operators } \hat{\phi}^\pm_{(n)} \in {\cal C}A_{2,0} 
\,|\, [\hat{J},\hat{\phi}^\pm_{(n)}]=\pm n\,\hat{\phi}^\pm_{(n)} \Big\},&\label{class+} \\[10pt] 
&\Psi^\pm_{(n+1/2)}:=\Big\{\hbox{All operators } \hat{\Psi}^\pm_{(n)} \in {\cal C}A_{2,1} 
\,|\, [\hat{J},\hat{\Psi}^\pm_{(n)}]=\pm (n+1/2)\,\hat{\Psi}^\pm_{(n)} \Big\}&\label{class-}
\end{eqnarray}
In other words, ${\cal C}A_{2}(P)$ is also $\mathbb Z$-graded.
A characteristic representant of any element in these sets can be chosen to be of the form
\begin{eqnarray}
\hat{\phi}^\pm_{(n)}\rightarrow \hat{\phi}^\pm_{\mu_1\mu_2 ...\mu_n}   , \qquad \hat{\Psi}^\pm_{(n+1/2)} \rightarrow\hat{\Psi}^\pm_{\mu_1\mu_2 ...\mu_n}{}^a \qquad n=0,1,2,\cdots\label{hatfield-2}
\end{eqnarray}
In fact, (\ref{class+}) and (\ref{class-}) form equivalence classes.
\end{enumerate}





\section{Conclusion}\label{concl}

We have considered infinite-component fields carrying Majorana's unitary representation of the Lorentz group.

\vspace{2mm}
On the one hand, imposing the Majorana equation, we have shown that it contains a massless superPoincar\'e multiplet with tensorial central charges. This supermultiplet is unusual because it carries continuous-spin representations of the Poincar\'e group. The possibility of a continuous-spin supermultiplet have been discussed already \cite{Brink:2002zx}, but without tensorial central extension. This supermultiplet is equivalently obtained as the continuous-spin limit of their massive solutions (\textit{cf.} \cite{Horvathy:2007pm}). 

On the other hand, imposing the Klein-Gordon equation, an explicit system is constructed which is supersymmetric. The corresponding realization of the superPoincar\'e algebra (\ref{Palg}) is rather exotic for three reasons: firstly, the ``translation'' operators (\ref{translgen}) are not equal to the momenta and instead act both on space-time and spinning coordinates; secondly, non-vanishing tensorial central charges are present; thirdly, there is no bound on the spin content, the supermultiplet contains an infinite tower of particles with equal mass $m$ for all spins $s\in{\mathbb N}/2\,$. Observe that in the case of the 
$\cal N$-extended supersymmetry algebra (without tensorial central charges), the difference between the maximal and minimal spin in any irreducible massive supermultiplet is at most equal to ${\cal N}/2$ (essentially because the supercharges are Grassmann odd). One could speculate that the exotic supersymmetric 
system constructed here should be related to the limit 
${\cal N} \rightarrow \infty$ of a supersymmetric theory.

\vspace{2mm}
Some comments on symmetry breaking are now in order. The massive (and tachyonic) solutions of the infinite-component Majorana equation are not supersymmetric but supersymmetry is approximately realized for large spins or small masses. It is shown by taking equation \eqref{susybr} and the Majorana mass-shell \eqref{MassSpectr}:
$$
[P\cdot\Gamma-M,Q^\pm_a]=\pm \frac{1}{2} Q^\pm_a \,m_J, \qquad m_J=\frac{M}{J+1/2},\qquad J=0,1/2,1,\ldots
$$ 
The supercharges $Q^\pm_a$ preserve the equation of motion only when the mass vanishes. Here the mass somehow acts as an order parameter (in analogy with Landau's terminology for its theory of second order phase transition); it is zero in the (super)symmetric phase and different from zero in the broken phase. Insisting on the analogy, the low-spin sector with strongly broken supersymmetry could correspond to the ordered/low-energy phase and, respectively, the higher-spin sector could correspond to its disordered/high-energy phase.
Thus, although the massless solutions of the Majorana equation are the only ones for which supersymmetry is exact, it might also be interesting to see if there exists some new spontaneous breaking of Poincar\'e supersymmetry where an infinite number of particles would acquire mass, except the continuous-spin supermultiplet which would appear as a Goldstone supermultiplet.

These ideas on symmetry breaking are also suggested from another point of view. 
Beyond the Planck scale, string theory is expected to reach an extremely symmetric phase with higher-spin gauge symmetries. Along this line of thinking, the infinite-dimensional higher-spin superalgebra on $AdS_4$ describing massless gauge fields of all spin (see \textit{e.g.} \cite{Vasiliev:1995dn} for some reviews) is expected to be spontaneously broken to its finite-dimensional subalgebra $\mathfrak{osp}(1|4)$ at low energies. Our degenerate massive spectrum carries a representation of the superalgebra $\mathfrak{osp}(1|4)$ and of its higher-spin extension. Therefore, the possibility arises that the dynamical origin of our massive supermultiplet might be some unusual flat limit of the non-Abelian higher-spin gauge theory on $AdS_4$.

\vspace{2mm}
\textit{A posteriori}, one might say that Majorana's work of 1932 even anticipated supersymmetry, confirming once again the premonitory character of his ideas. The fact that the spectrum contains only positive-energy solutions with fermions and bosons treated on equal ground was already an indication in favour of supersymmetry \cite{Zichichi:2006sf}. Another perspective on our observation is that if one supplements Majorana's equation by massless Klein-Gordon's equation, this set of field equations provides a supersymmetrization of Wigner's equations \cite{wi} describing the continuous-spin representation. It would be very interesting to investigate the precise relationship between these two descriptions, for instance by the possible link between the internal four-vector $\xi^\mu$ appearing in Wigner's equation and the infinite-dimensional ``Dirac matrices'' $\Gamma_\mu$ appearing in Majorana's equation (\ref{Diraceq}). 

Finally, it would be satisfactory to obtain the above-mentioned relativistic wave equations from an underlying particle/string/brane model via first quantization. For instance, single linear Regge trajectories have been obtained from various models: ``rigid'' string, ``composite'' particle, ``discrete'' string, \textit{etc} (see \textit{e.g.} \cite{Barducci:1976ge} and refs therein). As another example, continuous spin representations arise from several ``higher-order geometrical'' models generalizing \cite{Plyushchay:1990cu}
(see \textit{e.g.} \cite{Mourad:2006xk} and refs therein).
Moreover, the appearance of twistors in the Majorana representation and of tensorial central charges in the supersymmetry algebra are very reminiscent of the superparticle models \cite{Bandos:1999qf,Plyushchay:2003gv} and their developments. In these models, superparticles tensorial central charges have different physical meaning than that of superbranes. They correspond to spin degrees of freedom of the superparticles, while it is well known that brane
tensorial charges are similar to electric-magnetic charges. It would be interesting to find the right physical interpretation of the tensorial central charges appearing in the (continuous-spin and massive higher-spin) supermultiplets constructed in the present work. For this purpose, it is suggestive to look at its two dimensional internal space (see eqs. (\ref{DFPbos},\ref{DFPfer}) and comments below).

\vspace{2mm}
To conclude, the work initiated by Majorana could provide a useful toy model for testing various ideas on higher-spin spontaneous supersymmetry breaking.

\section*{Acknowledgments}

We thank N. Boulanger, M. Plyushchay and P. Sundell for valuable discussions. M.V. thanks to CNRS for postdoctoral grant (contract number 87366).

\appendix

\section{The Weyl algebra $A_2$ and some subalgebras} \label{app}
The complex Weyl algebra $A_2$ is generated by the $q_i$ and $\eta_j$
satisfying \eqref{A2}. If we consider $A_2^{\mathbb R}$,
 the real form of $A_2$, generated by 
the usual harmonic oscillators 
{
\beqa
\label{HO}
a^\pm_i=\frac{1}{\sqrt{2}}(q_i\mp \eta_i), \qquad i=1,2.
\eeqa
}
\noi
it admits a unitary representation given by 
{
\beqa
\label{A2-rep}
{\cal M}= \left\{\left|\,n_1,n_2\right> = \frac{(a_1^+)^{n_1}}{\sqrt{n_1 !}}
 \frac{(a_2^+)^{n_2}}{\sqrt{n_2 !}} \left|\,0\right>,
n_1, n_2 \in \mathbb N\right\},
\eeqa
}
\noi
with { the vacuum $ \left|\,0\right>$} being defined by  $ a^-_i \left|\,0\right>=0$.
Furthermore $A_2$ (or $A_2^{\mathbb R}$)
is a $\mathbb Z_2-$graded associative
algebra (graded by the map ${\cal R}$ (see \eqref{RR'}) and we have
$ A_2^\mathbb R = A^{\mathbb R}_{2,0} \oplus A^{\mathbb R}_{2,1},$  such that 
$A^{\mathbb R}_{2,i} A^{\mathbb R}_{2,j} \subseteq A^{\mathbb R}_{2,i+j}$
(modulo two).  Consequently the {  vector space $A_2^\mathbb R$ endowed with} the bracket
{ $[b_1,b_2]_\pm= b_1 b_2 -(-1)^{|b_1| |b_2|} b_2 b_1$ where $b_i$ are homogeneous
elements and $|b_i|$} their {  ${\mathbb Z}_2$-grading,} naturally inherits a structure of Lie superalgebra.
Of course, similar results hold for { the complex Weyl algebra} $A_2$.
It is known that various (super)algebras can be embedded in the Weyl algebra
$A_2^\mathbb R$
(the Lie subalgebra of the Weyl algebra $A_1$ have been classified
in \cite{rs}). Since the representation
\eqref{A2-rep} is unitary the corresponding representations 
will be automatically
unitary provided that the various generators are Hermitian combinations
of the oscillator \eqref{HO}.

\subsection{The $\mathfrak{osp}(1|4)$ Lie  superalgebra}
We construct now explicitly the $\mathfrak{osp}(1|4)$ algebra introduced in 
\eqref{so32g}. Its fermionic part is generated by the harmonic oscillators \eqref{HO}.
To construct its bosonic part, we
consider the Dirac matrices in the Majorana representation

\begin{equation}\label{gammas}
{\small
\gamma_0=\left(%
\begin{array}{cc}
  0 & -\sigma^0 \\
  \sigma^0 & 0 \\
\end{array}%
\right),\quad
\gamma_1=\left(%
\begin{array}{cc}
  0 & \sigma^0 \\
  \sigma^0 & 0 \\
\end{array}%
\right)},\quad {\small
\gamma_2=\left(%
\begin{array}{cc}
  \sigma^3 & 0 \\
  0 & -\sigma^3 \\
\end{array}\right),\quad
\gamma_3=\left(%
\begin{array}{cc}
  -\sigma^1 & 0 \\
  0 & \sigma^1 \\
\end{array}%
\right),}
\end{equation}

\noi
and set $\gamma_{\mu \nu}=-\frac{i}{4}[\gamma_\mu,\gamma_\nu]$. Using \eqref{so32g} we 
obtain

{
\beqa
\label{sp4}
\begin{array}{ll}
S_{12}=\frac12\left(N_1 -N_2\right),&
S_{23}=\frac{i}{2}\left(-a_2^+ a_1^- + a_1^+ a_2^-\right),\\
S_{31}=\frac12\left(a_2^+ a_1^- + a_1^+ a_2^-\right), &
S_{01}=\frac{i}{4}\left(-a_1^-{}^2+a_1^+{}^2-a_2^-{}^2+a_2^+{}^2\right),\\
S_{02}=\frac{1}{4}\left(a_1^-{}^2+a_1^+{}^2-a_2^-{}^2-a_2^+{}^2\right),&
S_{03}=\frac12\left(-a_1^- a_2^- -a_1^+ a_2^+\right), \\
S_{40}=\Gamma_0=\frac12\left(N_1 + N_2 +1\right),&
S_{41}=\Gamma_1=\frac14\left(a_1^-{}^2+a_1^+{}^2 + a_2^-{}^2+a_2^+{}^2\right),\\
S_{42}=\Gamma_2=\frac{i}{4}\left(a_1^-{}^2-a_1^+{}^2 - 
a_2^-{}^2+a_2^+{}^2\right),&
S_{43}=\Gamma_3=\frac{i}{2}\left(-a_1^- a_2^- + a_1^+ a_2^+\right),\\
\end{array}
\eeqa
}

\noi
it can be checked explicitly that the generators \eqref{sp4} are Hermitian.
Furthermore, since the odd part of the 
$\mathfrak{osp}(1|4)$ algebra is generated by the 
 usual harmonic oscillators  
the representations are automatically infinite-dimensional
and corresponds to \eqref{A2-rep}.
This is rather different to the usual supersymmetric theories, where the odd part
of the algebra is generated by Grassmann variables and consequently a supermultiplet
contains always a finite number of components.

We now identify the module space $\cal{M}$ in the decomposition 
$\mathfrak{so}(2,3) \subset A_2^\mathbb R$. 
We firstly identify the Cartan subalgebra, and the corresponding reduction of 
$\mathfrak{so}(2,3)$ under a rank-two subalgebra. 
Here we are interested in the reduction
 with respect to the maximal compact subalgebra 
$\mathfrak{so}(2)\oplus \mathfrak{so}(3)$. This algebra,  
 is also taken in account for the classification of particles on $AdS_4$ space {\cite{Nicolai:1984hb}}.
The  $\mathfrak{so}(2)\oplus \mathfrak{so}(3)$ subalgebra is generated by $\Gamma_0$ and 
$S_i:=S_{jk}$ ($i,j,k$  are in cyclic order). Its Cartan subalgebra is chosen to be 
formed by $\Gamma_0$ and $S_3$. It is interesting to observe that
 $\Gamma_0$ is up to a factor $1/2$  the Hamiltonian of a harmonic oscillator on the plane
(see \eqref{sp4}).
In this representation $\Gamma_0$ and the Casimir operator
of $\mathfrak{so}(3)$  are related

\begin{eqnarray}
\label{gamma0}
\Gamma_0=\hat{S}+\frac{1}{2},\qquad \vec S\cdot \vec S=\hat{S}(\hat{S}+1),
\qquad \hbox{where} \qquad \hat{S}=\frac{N_1+N_2}{2}. \label{u1xsu2}
\end{eqnarray}

\noi
Hence, the eigenvalues of the energy and the spin are integer or 
half-odd-integer numbers. However  this representation is remarkable
in the sense that when the eigenvalue of $\Gamma_0$ is an integer number,
the spin is an half-odd-integer number and conversely.
To identify precisely the representation of 
$\mathfrak{so}(2,3)$ we are dealing with, we introduce the
root operators
$E^{\epsilon,\epsilon'}=\frac{1}{\sqrt{2}}(M_1^{\epsilon'}
+ i \epsilon  M_2^{\epsilon'}), E^{0,\epsilon'}=M_3^{\epsilon'}$
(with $M_i^+=iS_{0i}-S_{4i}$ and $M_i^-=iS_{0i}+S_{4i}$)
 where $\epsilon,
\epsilon'=\pm$ are the eigenvalues of $S_{12}$ and $\Gamma_0$ respectively.
In particular we have
{
\beqa
\label{cartan}
E^{+,-}=(a_2^-)^2,\ \  E^{-,-}=(a_1^-)^2, \ \  E^{0,-}=-ia_1^- a_2^-.
\eeqa
}
\noi
The lowest weights of the vector modules are annihilated by 
$E^{+,-},E^{-,-}$ and $E^{0,-}$ 
and  are labelled by the eigenvalues of $\hat{S}$. 
Thus the triad $(\Gamma_0,\hat{S},S_3)$ takes the eigenvalues,
\begin{eqnarray}
\hbox{Rac}:\qquad (\Gamma_0,\hat{S},s_3){\cal M}_+&=&(S+1/2,S,s_3){\cal M}_+, 
\qquad S=0,1,2,3,\dots \label {Rac}\\[10pt]
\hbox{Di }:\qquad(\Gamma_0,\hat{S},s_3){\cal M}_-&=&(S+1/2,S,s_3){\cal M}_-, 
\qquad S=1/2,3/2,5/2,\dots\label{Di}
\end{eqnarray}
and $s_3=-S,-S+1,...,S-1,S.$ The lowest eigenvalues of the energy and spin operators
 $(\Gamma_0,\hat{S})$ are respectively $(1/2,0)$ and $(1,1/2)$, hence, every module 
corresponds to the Rac and Di representations
and are generically denoted ${\cal D}(1/2,0)$ and ${\cal D}(1,1/2)$ respectively. 
Taking in account (\ref{RR'}) and (\ref{u1xsu2}), 
\begin{equation}\label{phase}
{\cal R}=(-1)^{2\hat{S}}=\exp(i2\pi\hat{S}),
\end{equation}
the reflection operator becomes a {\it statistical phase}, i.e. a representation of the 
fundamental group {  $\mathbb{Z}_2$ of the rotation group.}

\subsection{The $\mathfrak{so}(3,1)$  algebra}

We would like now to show that  the $\mathfrak{so}(2,3)-$modules
 ${\cal D}(1/2,0)$ and ${\cal D}(1,1/2)$
remain irreducible {  under the restriction to  $\mathfrak{so}(1,3)$.}
First of all as a direct consequence of \eqref{sp4} and \eqref{gamma0}
we have that in the embedding $\mathfrak{so}(2) \oplus
\mathfrak{so}(3) \subset \mathfrak{so}(2,3)$

\beqa
\label{massive}
{\cal D}(1/2,0)&=& \sum \limits_{s \in \mathbb N} {\cal D}_s, \nonumber \\
{\cal D}(1,1/2)&=& \sum \limits_{s \in \mathbb N} {\cal D}_{\frac12 + s},
\eeqa

\noi
where ${\cal D}_s$ is the spin-$s$ { irreducible} representation of $\mathfrak{so}(3)$.
Furthermore, using \eqref{sp4} it is not difficult to see that
the boost operators ($K_i=S_{0i}$) maps states of different spin

$$
\xymatrix{
{\cal D}_0 \ar[r]^{K }& {\cal D}_1  \ar[r]^{K} & \cdots  \ar[r]^{K }& {\cal D}_n \ar[r]^{K }&
\cdots , \\
{\cal D}_{\frac12} \ar[r]^{K }& {\cal D}_{\frac32}  \ar[r]^{K} & \cdots  \ar[r]^{K }& 
{\cal D}_{\frac{2n+1}{2}} \ar[r]^{K }&
\cdots , \\
}
$$

\noi
and thus ${\cal D}(1/2,0)$ and ${\cal D}(1,1/2)$ remain irreducible {under the Lorentz group}.
 Recall  that the unitary representations of
$\mathfrak{so}(1,3)$ are usually denoted \cite{Casalbuoni:2006fa,st,ggv} by $[\ell_0,\ell_1]$ 
with either $\ell_1=i\sigma/2 \in i \mathbb R$ and $\ell_0 \in \frac12 \mathbb N$
(principal series) 
or $\ell_0=0$ and $0<\ell_1 \le 1$ (complementary series). 
The spin content of the representation
is $\ell_0, \ell_0 + 1, \cdots, \ell_0+n,\cdots$
and the Casimir operators are given 
{  by $${\cal C}_2\big(\mathfrak{so}(1,3)\big)=\frac12\, S_{\mu \nu} S^{\mu \nu}=
\ell_0^2+\ell_1^2-1$$ and $${\cal C}^\prime_2\big(\mathfrak{so}(1,3)\big)=\frac14\, \epsilon_{\mu \nu \rho \sigma} S^{\mu \nu}
S^{\rho \sigma} = -i \ell_0 \ell_1\,.$$ From}
\eqref{Casimir-Poin}, we have
 $\ell_0 \ell_1=0$ and $\ell_0^2+\ell_1^2=\frac14$.
Thus, in the embedding $\mathfrak{so}(1,3) \subset \mathfrak{so}(2,3)$, 
we have ${\cal D}(1/2,0) = [0,1/2]$ (complementary series), ${\cal D}(1,1/2)=[1/2,0]$
(principal series).
This can also be obtained simply by computing the matrix element of the
boost operator using \eqref{sp4} and comparing with 
\cite{Casalbuoni:2006fa,st}.

\subsection{The algebra associated to tachyons and massless particles}\label{tachyons}

It is possible also to classify the representations of $\mathfrak{so}(2,3)$ 
employing other 
stabilizer algebras as we now show. It could be associated to some exotic 
particles in $AdS_4$ or, as we 
shall see, to the massless and tachyonic particles {  on ${\mathbb R}^{1,3}$ (in Majorana's theory).} These subalgebras are 
respectively $\mathbb R \oplus \mathfrak{iso}(2)$, 
$\mathfrak{so}(1,1)\oplus \mathfrak{so}(1,2)$.
The results we recall here are given in \cite{Sudarshan:1970ss}. 
In table \ref{tab-ads} we summarize it for the reduction under the several 
subgroup.
\begin{table}[h]
\caption{}
\begin{center}
\begin{tabular}{|c|c|c|c|}
 \hline
$\mathfrak{so}(3,2)$ Little algebra & Invariant vector & Eigenvalues \\ 
\hline\hline
$\mathfrak{so}(2)\oplus \mathfrak{so}(3):\{\Gamma_0\}\oplus \{S_1,S_2,S_3\}$ 
& $\Gamma_0$ & $1/2,1,3/2,...$      \\\hline
$\mathfrak{iso}(2):S_3,\pi_1,\pi_2$ &  $\Gamma_0+\Gamma_3$    
& $0<\varepsilon <\infty$ \\\hline
$\mathfrak{so}(1,1)\oplus\mathfrak{so}(1,2):\{\Gamma_3\}\oplus\{S_3,S_{10},
S_{20}\}$ &  $\Gamma_3$  &$-\infty<\sigma <\infty$ \\\hline
\end{tabular}
\end{center}
\label{tab-ads}
\end{table}

We now study the decomposition under $\mathfrak{so}(1,1) 
\oplus \mathfrak{so}(1,2)$.
This embedding corresponds to the little algebra of the Poincar\'e algebra for
tachyonic  representations $P_\mu P^\mu = \ell^2>0$.
We consider the standard vector $P^\mu=(0,0,0,\ell)$ 
the little algebra $\mathfrak{so}(1,2) \oplus \mathfrak{so}(1,1)$ is then
generated by 
$\left\{S_{12},S_{20},S_{10}\right\} \oplus \left\{\Gamma_3\right\}$.
The Casimir operator of $\mathfrak{so}(1,2)$ is given by

$$
{\cal C}_2\Big(\mathfrak{so}(1,2)\Big)=J_{12}^2-J_{20}^2-J_{10}^2,
$$

\noi
and a direct computation using \eqref{sp4} gives

\beqa
\label{Q}
{\cal C}_2\Big(\mathfrak{so}(1,2)\Big)=\frac14-\Gamma_3^2.
\eeqa

\noi
This means that the representations of $\mathfrak{so}(2,1)$ can be 
characterised by means of the eigenvalues of $\Gamma_3$.
Unitary representation of $\mathfrak{so}(1,2)$ are either bounded from
below/above (discrete series) or unbounded from below and above
(continuous principal or continuous supplementary series).
The continuous principal series are characterised by a continuous
value of the Casimir operator ${\cal C}_2=\Phi(\Phi+1)$ with $\Phi=-\frac12 + i \rho,
\ \rho \in \mathbb R$ {\it i.e.} ${\cal C}_2=\frac14-\rho^2$. 
Since $\Gamma_3$ is a hermitian operator, its eigenvalue are reals.
Looking to \eqref{Q} means that for any eigenvalue $\sigma \in \mathbb R$
of $\Gamma_3$ we have a continuous principal series 
where ${\cal C}_2= \frac14 -\sigma^2$.
Indeed,  denote $\left|n_1,n_2\right>= 
\lb\ \frac{n_1+n_2}{2},\frac{n_1-n_2}{2}\ \rb$ an eigenvector of $\vec S\cdot\vec S$
and of $S_3$ with eigenvalue $j=(n_1+n_2)/2$ and $m=(n_1-n_2)/2$ respectively.
Introduce then for $n_2 \ge n_1$ (a similar case holds when $n_1 < n_2$) 

\beqa
\label{sigma}
\left|\sigma, m\right>= \sum \limits_{j \ge m}^{+\infty}
A_j^{\sigma ,m}\lb \  j,m\ \rb,
\eeqa

\noi
which is an eigenvector of $J=S_{12}=\frac12 (N_1-N_2)$ with eigenvalue
$m$ and impose $\Gamma_3\left|\sigma, m\right>=\sigma 
\left|\sigma, m\right>$.
This leads to an inductive relation for the coefficients $A_j^{\sigma,m}$
whose solution is \cite{Sudarshan:1970ss}

\beqa
\label{A}
A_j^{\sigma,m}=\frac{1}{\sqrt{2 \pi}} \frac{1}{\Gamma(2m+1)}
\sqrt{\frac{\Gamma(j+m+1)}{j-m+1}} 2^{m+\frac12} e^{-i\frac{\pi}{2}(m-j)}
F(m-j,m+\frac12 - \sigma,2m+1,2),
\eeqa

\noi
on account to  the identity upon the
hypergeometric functions  $F$ \cite{Bat}

$$
\big((b-a)z-c+2a\big)F(a,b,c,z)=
(a-c)F(a-1,b,c,z) +(1-z)a F(a+1,b,c,z).
$$

\noi One can show that these eigenstates are not normalisable and satisfy

$$
\left<\sigma',m'|\sigma,m\right>=\delta_{mm'} \delta(\sigma-\sigma').
$$

Now acting with the operators $K_\pm=S_{02} \pm i S_{01}$ upon \eqref{sigma},
one obtains a representation unbounded from bellow and from above
corresponding the the principal series ${\cal D}_p(\frac14-\sigma^2,0)$ if
$m$ is an integer number or ${\cal D}_p(\frac14-\sigma^2,\frac12)$ if
$m$ is a half-odd-integer number.
Consequently, if in the tachyonic case, where $P_\mu P^\mu =\ell^2$,
one solve the Majorana equation in the standard frame, we obtain the
principale series above with $\sigma=M/\ell$ as a solution.

\bigskip
We know study the
decomposition $\mathbb R \oplus \mathfrak{iso}(2)  \subset\mathfrak{so}(3,2)$.
This embedding corresponds to the little algebra of the Poincar\'e algebra for
massless  representations $P_\mu P^\mu =0$. In this case in the standard 
frame we have 
$P^\mu=({  E},0,0,{ E})$  and  
the little algebra $\mathfrak{iso}(2)$ is generated by
$J=J_{12}$, and $\pi_1= J_{01}+J_{31}, \pi_2=J_{02} + J_{32}$.
We thus consider the algebra $\mathbb R \oplus \mathfrak{iso}(2)$ generated 
by 
$\{J_{12},\pi_1,\pi_2\}$ and $\{ \Gamma_0 + \Gamma_3\}$. 
The Casimir operator of $\mathfrak{iso}(2)$ is given by 
$\pi^2=\pi_1^2+\pi_2^2$.
Using \eqref{sp4}, a direct computation leads to

$$
\pi^2=(\Gamma_0+\Gamma_3)^2.
$$  

\noi
Unitary representation of $\mathfrak{iso}(2)$ are infinite dimensional and
 defined
by ${\cal D}_0(p)=\left\{ \left|p,m\right>, m \in \mathbb Z \right\}$ or
${\cal D}_{\frac12}(p)=\left\{ \left|p,m+\frac12\right>, m \in \mathbb Z \right\}$

\beqa
&\pi^2  \left|p,m\right>=p^2   \left|p,m\right>,  
J  \left|p,m\right>= m  \left|p,m\right>, \nonumber \\ 
&\pi_+  \left|p,m\right>= -ip \left|p,m+1\right>,   
\pi_-  \left|p,m\right>= ip \left|p,m-1\right>,
\eeqa

\noi
with $\pi_\pm=\pi_1\pm i \pi_2$ and 
$m \in \mathbb Z$ or $\mathbb Z + \frac12$. 
Such representations are called the continuous spin representations.
As before, one is able for any value of $\lambda \in \mathbb R^*$,
to find an eigenvector of $\Gamma_0 + \Gamma_3$ with an eigenvalue
$\lambda$ leading thus to the continuous spin representation
$D_0(\epsilon)$ and $D_\frac12(\epsilon)$ 
with $\epsilon = \lambda^2>0$ \cite{Sudarshan:1970ss}. Furthermore, if
one solves the Majorana equation in the standard frame, one obtains
the two continuous spin representation above with $\epsilon = M$.


\end{document}